\documentclass[11pt]{article}

\usepackage[round]{natbib}
\usepackage{amssymb, amsthm, amsmath, graphicx, float, multirow, mathrsfs, amsfonts, enumitem}
\usepackage{subcaption}
\usepackage{bm}
\usepackage{comment}

\theoremstyle{plain}

\newtheorem{theorem}{Theorem}
\newtheorem{lemma}{Lemma}

\newtheorem{proposition}{Proposition}

\theoremstyle{definition}

\newtheorem{definition}{Definition}

\theoremstyle{remark}
\newtheorem{remark}{Remark}

\newcommand{\taubar}{\bar{\tau}}
\newcommand{\taubarhat}{\widehat{\bar{\tau}}}
\newcommand{\taubarhatHT}{\widehat{\bar{\tau}}^{\text{HT}}}
\newcommand{\YbarhatHTz}{\widehat{\bar{Y}}^{\text{HT}}(z)}
\newcommand{\Ybarz}{\bar{Y}(z)}
\newcommand{\Ybarhatz}{\widehat{\bar{Y}}(z)}

\newcommand{\Ybarhathz}{\widehat{\bar{Y}}_h(z)}
\newcommand{\Ybarhatzstar}{\widehat{\bar{Y}}(z^*)}

\usepackage{color}

\DeclareMathSymbol{\Y}{\mathbin}{AMSb}{"59}
\DeclareMathSymbol{\bS}{\mathbin}{AMSb}{"53}

\usepackage{authblk}

\title{Causal Inference in Rebuilding and Extending the Recondite Bridge between Finite Population Sampling and Experimental Design}
\author[*]{Rahul Mukerjee}
\author[**]{Tirthankar Dasgupta}
\author[**]{Donald B. Rubin}
\affil[*]{Indian Institute of Management Calcutta}
\affil[**]{Department of Statistics, Harvard University}

\begin{document}

\date{}

\maketitle

\small

\begin{abstract}
This article considers causal inference for treatment contrasts from a randomized experiment using potential outcomes in a finite population setting. Adopting a Neymanian repeated sampling approach that integrates such causal inference with finite population survey sampling, an inferential framework is developed for \emph{general} mechanisms of assigning experimental units to multiple treatments. This framework extends classical methods by allowing the possibility of randomization restrictions and unequal replications. Novel conditions that are ``milder'' than strict additivity of treatment effects, yet permit unbiased estimation of the finite population sampling variance of any treatment contrast estimator, are derived. The consequences of departures from such conditions are also studied under the criterion of minimax bias, and a new justification for using the Neymanian conservative sampling variance estimator in experiments is provided. The proposed approach can readily be extended to the case of treatments with a general factorial structure.
\end{abstract}


\normalsize 
\section{Introduction}

Finite population sampling and experimental design were two of most researched topics in statistics during the first part of the twentieth century, with both being rooted in the fundamental statistical principle that inferences should be based only on the randomization distribution actually used, a perspective explicitly first described in \cite{Fisher1925}. Ideas from finite population survey sampling were used in the development of the theory of design and analysis of randomized experiments. A pioneering effort along these lines was the seminal work of \cite{Neyman:1923}, in which he introduced the concept of \emph{potential outcomes} and used it when deriving the sampling, or randomization, distribution of the estimated average treatment effect. During the next few decades, several researchers, e.g., \cite{kempthorne1955randomization}, \cite{wilk1955randomization}, \citet{wilk1956some,wilk1957non} worked along similar lines. As randomization became more complex, leading to development of designs with randomization restrictions (e.g., block designs, Latin square designs, split-plot designs), derivation of sampling distributions of estimators of treatment effects became increasingly difficult. Plagued by the lack of computational resources to check the correctness of analytical results, these derivations were sometimes incorrect $-$ see \cite{Sabbaghi2014} on \cite{Neyman:1935}.

Development of experimental design methodology in the later part of the twentieth century took a path often dominated by its application to manufacturing and the connection to finite population survey sampling became recondite at best. Because in a controlled manufacturing environment, superpopulation inference is typically of more interest than its finite population counterpart, and a finite population of experimental units can often be viewed as a random sample from a larger population (defined by controllable manufacturing conditions), model-based inference with hypothetical parameters (e.g., the normal linear model) gradually gained popularity and dominated textbooks such as \cite{HamadaWu2009}.

The potential outcomes framework made a resurgence through its application to causal inference from randomized experiments and observational studies in the social, behavioral, and biomedical (SBB) sciences. Conception and development of this perspective of causal inference can be attributed to the formalization of Neyman's \citep{Neyman:1923} notation and results by \citet{Rubin:1974, Rubin:1975, Rubin:1977} in what has been called the ``Rubin Causal Model'' or RCM \citep{Holland1986}. However, commonly SBB experiments (see \cite{Imbens2015}) involve only a single factor at two levels (e.g., treatment-control studies) and there exists limited literature on the use of the potential outcomes framework to draw inference from experiments with multiple treatments. Such SBB experiments involving multiple treatments, often with a factorial structure, appear to be gaining popularity recently (see, for example, \cite{chakraborty2009}, \cite{collins2009}).

Recently, \cite{Dasgupta:2015} considered the problem of drawing causal inference from balanced $2^K$ factorial experiments using the potential outcomes framework. Specifically, for a finite population of experimental units, they studied sampling distributions of estimators of factorial effects under a balanced completely randomized treatment assignment mechanism and investigated how the sampling variances of these estimators were affected by departures from the common ``Neymanian assumption'' of strict additivity of treatment effects (which means all units have exactly the same treatment effect). These results were extended to the case of unbalanced $2^K$ experiments by \cite{Lu:2016}. However, completely randomized SBB experiments with exactly two levels for each factor often cannot be conducted. To facilitate application to SBB experiments, it is important to consider assignment mechanisms and treatment structures that are more general than completely randomized $2^K$ experiments.

\medskip

This article attempts to reconstruct the nearly broken bridge between finite population survey sampling and experimental design with the aim of developing a framework for causal inference for treatment contrasts from randomized experiments with a \emph{general assignment mechanism} with in a finite population setting. The assignment mechanisms considered allow the possibility of randomization restrictions, such as stratification or split-plot structure, and unequal replications. A new result shows that unbiased estimation of the sampling variance of any treatment contrast estimator is possible under conditions milder than standard (e.g., without strict additivity). The consequences of departures from such conditions are also studied, and a new justification behind using the so-called conservative ``Neymanian variance estimator'' of the estimated treatment contrast is provided by considering minimax bias. The proposed approach applies readily to the situation where the treatments have a \emph{general} factorial structure with $K$ factors, each with a possibly different number of levels.

The article is organized as follows: Section \ref{s:PO} gives a brief overview of potential outcomes and associated estimands (quantities of interest) at the unit-level and the population-level. In Section \ref{s:assignment}, a general assignment mechanism is introduced and illustrated using three examples that explain its connection to finite population sampling theory. Section \ref{s:Neyman} develops the Neymanian inference procedure under the general assignment mechanism and presents results that identify conditions, milder than strict additivity, permitting unbiased estimation of the sampling variances (and covariances) of linear estimators of treatment contrasts. These conditions are examined in Section \ref{s:conditions}, and two examples are used to illustrate their applicability. Section \ref{s:bias} explores the effect of departures from additivity (strict and mild) on the bias of the estimators and identifies the Neymanian variance estimator as a minimax solution. Section \ref{s:factorial} shows how the proposed framework can be applied to factorial experiments, and Section \ref{s:simulations} illustrates the proposed approach using some simulated numerical examples. Section \ref{s:discussion} presents a discussion and indicates future research.

\section{Potential outcomes, unit-level and population-level treatment contrasts} \label{s:PO}

Consider a randomized experiment that involves allocation of a set $Z$ of treatments to a population $\{1, \ldots, N\}$ of $N$ experimental units such that each unit is assigned to one of the $|Z|$ treatments ($|Z|$ denotes the cardinality of the set $Z$). For each $z \in Z$ and  $i= 1, \ldots,N$, let $Y_i(z)$ denote the potential outcome of unit $i$ if assigned to treatment $z$. A unit-level treatment contrast for unit $i$  is of the form:
\begin{equation}
\tau_i = \sum_{z \in Z} g(z) Y_i(z), \label{eq:unitlevelcontrast}
\end{equation}
where $g(z)$, $z \in Z$, are known, not all zeros, and sum to zero. Such a contrast is a unit-level causal estimand under the RCM that compares the potential outcomes of a unit when exposed to different treatments. Let $\bar{Y}(z) = \sum_{i=1}^N Y_i(z) / N, z \in Z$ denote the mean of the $N$ potential outcomes for treatment $z$ across all units.  Then the mean of the unit-level estimands $\tau_1, \ldots, \tau_N$, namely,
\begin{equation}
\bar{\tau} = (\tau_1 + \ldots + \tau_N)/N = \sum_{z \in Z} g(z) \bar{Y}(z), \label{eq:poplevelcontrast}
\end{equation}
defines a population-level treatment contrast, which is a causal estimand for the finite population of units. Our interest lies in Neymanian inference \citep{Rubin:1990} for treatment contrasts, like $\bar{\tau}$,  on the basis of the outcomes observed from the experiment. If the treatments have a factorial structure, then treatment contrasts representing the factorial main effects and interactions are typically of interest. Such contrasts will be defined in Section \ref{s:factorial}.

\section{A general assignment mechanism for randomized experiments} \label{s:assignment}

An assignment mechanism is the rule that assigns experimental units to treatments; see \cite{Imbens2015} for a detailed discussion of assignment mechanisms. Here we consider a general assignment mechanism for randomized experiments that is inspired by a general sampling design for a finite population (e.g., \cite{HHM:1953}, \cite{Kish:1965}, \cite{Cochran:1977}). Let $T =\{T(z): z \in Z\}$ be any partition of the population $\{1, \ldots,N\}$ of experimental units into $|Z|$ nonempty disjoint subsets $T(z), z \in Z$. We consider a general mechanism for assigning units to treatments that involves the following two steps:
\begin{enumerate}
\item[(i)]  Select a partition of the $N$ units, $T =\{T(z): z \in Z \}$, where the partition has known probability $p(T) (\ge 0)$ of being selected, and
\item[(ii)] assign all the units in $T(z)$ to treatment $z$, for every $z \in Z$.
\end{enumerate}

\noindent In the remaining part of this paper, we shall refer to the above assignment mechanism as a \emph{general assignment mechanism} (GAM), which must satisfy $\sum_T p(T) = 1$, where $\sum_T$ denotes the sum over all such partitions $T$. In the three examples discussed in Section \ref{ss:examples}, all $T$ with $p(T) > 0$ have the same $p(T)$, but this is irrelevant to the theory developed here. Implicitly, all probabilities are conditional in the sense that they treat the values of all potential outcomes and blocking factors as fixed.

Under a GAM, the observed outcome for unit $i$ can be expressed as:
\begin{eqnarray*}
Y_i^{\textrm{obs}} = \sum_{z \in Z} W_i(z) Y_i(z),
\end{eqnarray*}
where $W_i(z)$ is a binary random variable satisfying:
\begin{eqnarray*}
Pr \left(W_i \left(z \right) = 1 \right) = E \left( W_i \left(z \right) \right) = Pr \left( i \in T \left(z \right) \right), z \in Z;
\end{eqnarray*}
that is, $W_i(z)$ takes value 1 if unit $i$ is assigned to treatment $z$ and zero otherwise.

\subsection{Three examples} \label{ss:examples}

\noindent \textbf{Example 1: Stratified random assignment}. Let the population $\{1, \ldots, N\}$ of units be partitioned into $H$ strata $\Omega_1, \ldots, \Omega_H$ with $|\Omega_h| = N_h \ (\ge 2)$ for $h = 1, \ldots, H$, and $\sum_{h=1}^H N_h = N$. Consider a randomization, that assigns $r_h(z)$ units in $\Omega_h$ to treatment $z$, where $r_h(z) (\ge 2)$, $z \in Z$, are fixed integers that sum to $N_h \ (h=1, \ldots, H)$ and all assignments within each stratum are equiprobable.

Such an assignment mechanism can be equivalently described as follows: for $h = 1, \ldots, H$, let $T_h(z)$ denote the set of $r_h(z)$ units assigned to treatment $z$ in stratum $\Omega_h$. Then $T(z) = T_1(z)\bigcup \ldots \bigcup T_H(z)$ is the set of $\sum_{h=1}^H r_h(z)$ units from all the strata assigned to treatment $z$, and $\{T(z): z \in Z \}$ is a partition of the $N$ units into $|Z|$ treatment groups. The assignment mechanism selects one among all such possible partitions with equal probability. In particular, if there is a single stratum, i.e., $H = 1$, then the assignment is completely randomized, as defined in textbooks in classical experimental design.
				
\bigskip

\noindent \textbf{Example 2: Split-plot assignment}. Assume that the treatments have a factorial structure, representing treatment combinations of two factors $F_1$ and $F_2$, each at two or more levels. Let $Z_k$ denote the set of levels of $F_k \ (k = 1,2)$, and let $z = z_1 z_2$ denote a treatment combination, where $z_k \in  Z_k$. In many practical situations, the levels of $F_1$ are more difficult to change from unit to unit than the levels of $F_2$. In such situations, it can be advantageous to employ a ``split-plot'' assignment of units to treatment combinations. Let $N = H N_0$, for two integers $H$ and $N_0$, both larger than one, and partition the population of $N$ units into $H$ disjoint sets $\Omega_1, \ldots, \Omega_H$, which are called whole-plots, each of which consists of $N_0$ units, called sub-plots nested within each whole-plot. Consider now a two-stage randomization of treatments, that assigns $r_1(z_1) \ (\ge 2)$ whole-plots to level $z_1$ of $F_1$, and then, within each whole-plot, assigns $r_2(z_2)$ sub-plots to level $z_2$ of $F_2$. Here, we suppose that, first, all assignments at each stage are equiprobable; second, that the fixed positive integers, $r_1(z_1)$, $z_1 \in  Z_1$, sum to $H$; and third, that the fixed positive integers $r_2(z_2)$, $z_2 \in Z_2$, sum to $N_0$. This is equivalent to selecting, with equal probability, one of the  partitions of the form $\{ T(z_1z_2): z_1z_2 \in Z \}$, where
\begin{equation}
T(z_1 z_2) = \bigcup_{h \in T_1(z_1)} T_{h2}(z_2), \label{eq:split_plot}
\end{equation}
for each $z_1z_2$. Here $T_1(z_1)$ represents the set of whole-plots assigned to level $z_1$ of factor $F_1$, so that for $z_1 \in Z_1$, the sets $T_1(z_1)$ are disjoint having union $\{1, \ldots, H\}$. The set $T_{h2}(z_2)$ represents the set of sub-plots within whole-plot $\Omega_h$ that are assigned to level $z_2$ of factor $F_2$. For each $h = 1, \ldots, H$, the sets $T_{h2}(z_2)$, $z_2 \in Z_2$, are also disjoint having union $\Omega_h$. Moreover, $|T_1(z_1)|= r_1(z_1)$ and $|T_{h2}(z_2)| = r_2(z_2)$. It is important to note that, although the whole-plots $\Omega_1, \ldots, \Omega_H$ are formally similar to the strata in Example 1, the assignment mehanism is now different because of the two-stage randomization involved.

\bigskip

\noindent \textbf{Example 3: Unicluster assignment.} Partition the population of $N$ units into $|Z|$ disjoint clusters $\Delta(1), \ldots, \Delta(|Z|)$. Analogous to unicluster designs in finite population sampling \citep{Sinha1991} which include systematic sampling, consider an assignment mechanism that assigns all units in a cluster to one of the treatments completely at random, that is, the treatments assigned to clusters $\Delta(1), \ldots, \Delta(|Z|)$ form a random permutation of $1, \ldots, |Z|$, all such permutations being equiprobable. This is equivalent to selecting, with equal probability, one of the $|Z|!$ partitions $\{T(z): z \in Z\}$, where each $T(z)$ equals one of the clusters. 		

\subsection{Assignment probabilities of first two orders} \label{ss:probability}

Recall that $W_i(z)$ denotes the binary treatment indicator that unit $i$ receives treatment $z$. The assignment probabilities of the first two orders are then given by $\pi_i(z) = E \left[ W_i\left(z\right) \right]$  and $\pi_{ii^*}(z,z^*) = E \left[ W_i\left(z\right) W_{i^*} \left(z^* \right) \right] $, where $i, i^* (\ne i)   = 1, \ldots, N$, and $z, z^* \in Z$. Thus, $\pi_i(z)$ is the probability that unit $i$ is assigned to treatment $z$. Similarly, $\pi_{ii^*}(z,z^*)$ is the probability that unit $i$ is assigned to treatment $z$ and unit $i^*$ is assigned to treatment $z^*$. Much of the theoretical developments under a GAM that will be developed in the next few sections depends on these quantities.

Recall that our interest lies in estimation of linear combinations (contrasts) of treatment means $\bar{Y}(z), z \in Z$. In finite population sampling \citep[e.g., see][]{Kish:1965} a necessary and sufficient condition to allow unbiased estimation of these treatment means is:
\begin{equation}
\pi_i(z) > 0, \;\, i=1, \ldots, N, \;\, z \in Z, \label{eq:NScondition}
\end{equation}
where these $\pi_i(z)$ are known. Thus the GAM will, hereafter, be assumed to satisfy condition (\ref{eq:NScondition}). Conditions involving $\pi_{ii^*}(z,z^*)$ will be imposed in Section \ref{s:Neyman} for estimation of sampling variance.

The assignment probabilities for the assignment mechanisms in the three examples introduced in Section \ref{ss:examples} are shown in the Appendix.

\section{Neymanian inference for treatment contrasts} \label{s:Neyman}

We now focus on our main objective, i.e., inferring the values of finite population treatment contrasts defined in (\ref{eq:poplevelcontrast}), i..e., Neymanian randomization-based inference for the finite population treatment contrasts, based on evaluation of expectations of statistics over the distribution induced by a GAM \citep{Rubin:2008}. From (\ref{eq:poplevelcontrast}), it is clear that an unbiased estimator of the finite population contrast  $\bar{\tau}$ is the obvious method of moments estimator
\begin{equation}
\widehat{\bar{\tau}} = \sum_{z \in Z} g(z) \widehat{\bar{Y}}(z), \label{eq:tauest}
\end{equation}
where $\widehat{\bar{Y}}(z)$ is an unbiased estimator of $\bar{Y}(z)$. In Section \ref{s:Neyman} we will compute the sampling properties of this estimator under a GAM. Specifically, in Section \ref{ss:LUE}, we will introduce linear unbiased estimators (LUE) of $\bar{Y}(z)$ from the finite population sampling literature, substitute the LUE in (\ref{eq:tauest}) to obtain an LUE of $\bar{\tau}$, and obtain an analytical expression for its sampling variance with respect to the distribution induced by the GAM. In the subsequent Sections \ref{ss:estNeyvar} and \ref{ss:estNeycov}, we discuss the estimation of sampling variance and covariance, respectively, and introduce a new approach.

\subsection{LUEs of treatment contrasts and their sampling variances} \label{ss:LUE}

For a GAM, let
\begin{equation}
\widehat{\bar{Y}}(z) = a(T,z) + \sum_{i \in T(z)} b_i(T,z) Y_i(z), \label{eq:LUE}
\end{equation}
be any LUE of the treatment mean ${\bar{Y}}(z), z \in Z$, where $a(T,z)$ and $b_i(T,z)$ are known quantities such that $a(T,z)$ depends on $z$ and the selected partition $T$, and $b_i(T,z)$ depends, also, on $i$. An example of such a LUE of ${\bar{Y}}(z)$ is the popular Horvitz-Thompson estimator \citep{HT:1952}, which is a kind of weighted average of the observed outcomes of all units exposed to treatment $z$, where the weight for each unit is the inverse of its first-order assignment probability, i.e., in (\ref{eq:LUE}), $a(T,z)=0$ and $b_i(T,z) = \left( N \pi_i\left(z \right) \right)^{-1}$. This estimator, henceforth denoted by $\widehat{\bar{Y}}^{\textrm{HT}}(z)$, is
\begin{equation}
\widehat{\bar{Y}}^{\textrm{HT}}(z) = \sum_{i \in T(z)} \frac{Y_i(z)}{N \pi_i(z)}. \label{eq:HT}
\end{equation}

Given a LUE $\widehat{\bar{Y}}(z)$ of the form (\ref{eq:LUE}), a LUE $\widehat{\bar{\tau}}$ of the estimand of interest, $\bar{\tau}$, can be obtained by substituting $\widehat{\bar{Y}}(z)$ from (\ref{eq:LUE}) into (\ref{eq:tauest}). The sampling variance of  $\widehat{\bar{\tau}}$ is
\small
\begin{equation}
var(\widehat{\bar{\tau}}) = E(\taubarhat^2) - \taubar^2 = \sum_{z \in Z} \sum_{z^* \in Z} g(z) g(z^*) E \left( \Ybarhatz \Ybarhatzstar \right) - \bar{\tau}^2. \label{eq:vartau}
\end{equation}
\normalsize

\noindent To calculate the first term on the right hand side of (\ref{eq:vartau}), we use the following lemma, that is a standard result in finite population sampling literature \citep{Chaudhuri2014}:

\begin{lemma} \label{lemma1}
If $\Ybarhatz$ is a LUE of $\Ybarz$ and is of the form (\ref{eq:LUE}), then
\begin{eqnarray}
 E \left( \Ybarhatz \Ybarhatzstar \right) &=& A(z, z^*) + \sum_{i=1}^N \left( A_i^{(1)}\left(z, z^*\right) Y_i(z) + A_i^{(2)} \left(z, z^* \right) Y_i(z^*) \right) \nonumber \\
 &+& \sum_{i=1}^N \sum_{i^*=1}^N B_{i i^*} (z, z^*) Y_i(z) Y_{i^*}(z^*), \label{eq:lemma1}
\end{eqnarray}
where $A(z, z^*)$, $A_i^{(1)}\left(z, z^*\right)$, $A_i^{(2)}\left(z, z^*\right)$ and $B_{i i^*} (z, z^*)$ are known constants that may depend on $z$ and $z^*$ as well as the respective subscripts. In particular,
\begin{equation}
B_{i i^*}(z, z^*) = \sum_{T: i \in T(z), i^* \in T(z^*)} p(T) b_i(T,z) b_{i^*}(T, z^*). \label{eq:Bij}
\end{equation}
\end{lemma}

\bigskip

In a GAM, because unit $i$ cannot be assigned to both $z$ and $z^*$, unit $i$ cannot be in $T(z)$ and $T(z^*)$. Therefore for $i = 1, \ldots, N$, the term $B_{ii}(z, z^*)$ in Lemma \ref{lemma1} vanishes whenever $z \ne z^*$. The following result can now be established using (\ref{eq:vartau}) and equation (\ref{eq:lemma1}) of Lemma \ref{lemma1}.

\begin{theorem} \label{thm1}
Under a GAM, the sampling variance of $\taubarhat$ is:
\small
\begin{eqnarray}
var(\taubarhat) &=& M + \sum_{z \in Z} \sum_{i=1}^N \left( M_i \left( z \right) Y_i \left( z \right) + M_{ii} \left(z \right) \left( Y_i \left( z \right) \right)^2 \right) \nonumber \\
&+& \sum_{z \in Z} \sum_{z^* \in Z} \sum_{i=1}^N \sum_{i^* (\ne i) = 1}^N M_{i i^*} (z, z^*) Y_i(z) Y_{i^*}(z^*) - \taubar^2, \label{eq:thm1varest}
\end{eqnarray}
where
\begin{eqnarray}
M = \sum_{z \in Z} \sum_{z^* \in Z} g(z) g(z^*) A(z, z^*), \  && \, M_i(z) = g(z) \sum_{z^* \in Z} g(z^*) \left( A_i^{(1)} \left(z, z^* \right) + A_i^{(2)} \left(z^*,z \right) \right), \nonumber \\
M_{ii}(z) = \left( g(z) \right)^2 B_{ii}(z,z), \ && \ M_{i i^*}(z, z^*) = g(z) g(z^*) B_{i i^*}(z, z^*), \label{eq:M}
\end{eqnarray}
for $z, z^* \in Z$ and $i, i^*(\ne i) = 1, \ldots, N$.
\end{theorem}
\normalsize

\bigskip

\begin{remark} \label{remark:Implications of Theorem 1}

Theorem \ref{thm1}, and the later results, generalize several results on the sampling variance of estimators of causal estimands available in the experimental design and causal inference literature. Consider the important special case of our formulation with two treatments, $Z = \{0,1\}$. Let the estimand of interest be the average treatment effect $\taubar = \bar{Y}(1) - \bar{Y}(0)$, so that $g(1) = 1$ and $g(0) = -1$. Consider the stratified assignment mechanism described in Example 1 of Section \ref{ss:examples} with $H=1$, i.e., a completely randomized assignment. Let $r(1) \ (\ge 2)$ and $r(0) \ (\ge 2)$ units be exposed to treatments 1 and 0 respectively, and consider the HT-estimator $\taubarhatHT = \sum_{i \in T(1)} Y_i(1) / r(1) - \sum_{i \in T(0)} Y_i(0) / r(0)$ as a LUE of $\taubar$. Then, after some non-trivial algebraic simplifications that are similar to results discussed later in Section \ref{ss:neat_examples}, the following expression for the sampling variance of $\taubarhatHT$ is obtained from (\ref{eq:thm1varest}) :
\begin{eqnarray}
var(\taubarhatHT) &=& \frac{\sum_{i=1}^N (Y_i(0) - \bar{Y}(0))^2}{(N-1)r(0)} + \frac{\sum_{i=1}^N(Y_i(1) - \bar{Y}(1))^2}{(N-1)r(1)} - \frac{\sum_{i=1}^N (\tau_i - \bar{\tau})^2}{N(N-1)} \nonumber \\
&=& \frac{S(0,0)}{r(0)} + \frac{S(1,1)}{r(1)} - \frac{S(\tau, \tau)}{N}, \label{eq:Ney_var}
\end{eqnarray}
where $S(z,z)$ is the variance of $Y_1(z), \ldots, Y_N(z)$ with divisor $N-1$, and $S(\tau, \tau)$ is the variance of unit-level treatment effects $\tau_1, \ldots, \tau_N$ also with divisor $N-1$.

The decomposition of this sampling variance into two components: $ S(0,0)/r(0) + S(1,1)/r(1) $ and $S(\tau, \tau)/N$ in (\ref{eq:Ney_var}) above, was originally developed by Neyman \citep{Neyman:1923}, and is a much studied result in causal inference literature (see, for example \cite{Rubin:1980}, \cite{Rubin:1990}, \cite{Gadbury:2001}, Ch. 7 of \cite{Imbens2015}, \cite{Dasgupta:2015}). This decomposition plays a crucial role in understanding the conditions under which $var(\taubarhatHT)$ can be unbiasedly estimated. Henceforth we refer to this decomposition as the ``Neymanian decomposition''.

\end{remark}

\subsection{Estimation of sampling variance} \label{ss:estNeyvar}

Even if (\ref{eq:NScondition}) holds and, in addition, the second-order assignment probabilities are all positive, $var(\taubarhat)$ does not admit an unbiased estimator because, by (\ref{eq:poplevelcontrast}), the last term, $\bar{\tau}^2$,  in (\ref{eq:thm1varest}) involves products like $Y_i(z)Y_i(z^*)$, and one can never observe both $Y_i(z)$ and $Y_i(z^*)$ for $z \ne z$*. This problem with estimation of the sampling variance was first noted by Neyman \citep{Neyman:1923}, which suggested that unbiased estimation of the sampling variance is only generally possible under the assumption of \emph{strict additivity}. Additivity and its impact on estimation of sampling variance of $\taubarhat$ has been studied by several authors since 1923. Recently, \cite{Dasgupta:2015} showed that for a completely randomized balanced $2^K$ factorial experiment, the Neymanian estimator of $\taubarhat$ has an upward bias that vanishes under strict additivity.
\begin{definition} \label{def:strict_add}
The potential outcomes $Y_i(z), i=1, \ldots, N, \  z \in Z,$ have \emph{strictly additive} treatment effects if for every $z, z^* \in Z$, $Y_i(z) - Y_i(z^*)$ is the same for $i=1, \ldots, N$, or equivalently, in matrix notation,
\begin{equation}
({\bm I}-N^{-1}{\bm J}) [{\bm Y}(z)-{\bm Y}(z^*)] = 0, \;\ 	z,z^* \in Z, \label{eq:strict_additivity}
\end{equation}
where ${\bm I}$ is the identity matrix and ${\bm J}$ is the matrix of all ones, both of order $N$, and ${\bm Y}(z)$ is the column vector with elements $Y_1(z), \ldots, Y_N(z)$ for each $z \in Z$.
\end{definition}

In our general formulation, which covers the one in \cite{Dasgupta:2015}, we explore unbiased estimation of $var(\taubarhat)$ under conditions that are milder than strict additivity. In what follows, any positive semidefinite (psd) matrix is assumed, as usual, to be also symmetric.

Let ${\mathbb Q}$ denote a class of psd matrices ${\bm Q}$ of order $N$ with $(i,i^*)$th element $q_{i i^*}$, such that the $q_{i i^*}$ are known constants and
\begin{equation}
{\bm Q}{\bm J} = {\bm 0}, \ \ \ \ q_{ii}=1/N^2, \ i= 1, \ldots, N. \label{eq:Q_cond}
\end{equation}
\noindent An example is:
\begin{equation}
{\bm Q}_{\text{strict}} =\left(N \left( N-1 \right) \right)^{-1} \left({\bm I}-N^{-1} {\bm J} \right). \label{eq:Q^*}
\end{equation}

\begin{remark} \label{remark: Qstrictconnection}
The connection between ${\bm Q}_{\text{strict}}$ and the Neymanian decomposition arises because the last term in (\ref{eq:Ney_var}) can be written as:
$$ \frac{S(\tau, \tau)}{N} = \frac{1}{N(N-1)} \sum_{i=1}^N (\tau_i - \taubar)^2 = {\bm \tau}^{\prime} {\bm Q}_{\text{strict}} {\bm \tau},$$
where ${\bm \tau} = (\tau_1, \ldots, \tau_N)^{\prime}$ is the vector of unit-level treatment contrasts. Also, from Definition \ref{def:strict_add}, the condition for strict additivity can be stated equivalently as
$${\bm Q}_{\text{strict}} [{\bm Y}(z)-{\bm Y}(z^*)] = 0, \;\ 	z,z^* \in Z. $$
\end{remark}

We now show how the class of matrices ${\mathbb Q}$ plays a crucial role in ``extracting'' an estimable component of $var(\taubarhat)$ and thus leads to a generalization of the Neymanian decomposition. Consider the quantity $M_{ii^*}(z, z^*)$ defined in the last line of (\ref{eq:M}) that appears in the third term of (\ref{eq:thm1varest}), and modify it:
\begin{equation}
\widetilde{M}_{ii^*}(z, z^*) = g(z) g(z^*) \left\{ B_{ii^*} (z, z^*) + q_{ii^*} - (1/N^2) \right\}, \label{eq:tildeM}
\end{equation}
where $q_{ii^*}$ is the $(i,i^*)$th element of ${\bm Q} \in {\mathbb Q}$. We now replace $M_{ii^*}(z, z^*)$ by $\widetilde{M}_{ii^*}(z, z^*)$ in (\ref{eq:thm1varest}), and drop the term $\bar{\tau}^2$ to define, for ${\bm Q} \in {\mathbb Q}$,
\small
\begin{eqnarray}
V_{\bm Q}(\taubarhat) = M + \sum_{z \in Z} \sum_{i=1}^N \left( M_i \left( z \right) Y_i \left( z \right) + M_{ii} \left(z \right) \left( Y_i \left( z \right) \right)^2 \right) + \sum_{z \in Z} \sum_{z^* \in Z} \sum_{i=1}^N \sum_{i^* (\ne i) = 1}^N \widetilde{M}_{ii^*} (z, z^*) Y_i(z) Y_{i^*}(z^*). \nonumber \\ \label{eq:upperbound}
\end{eqnarray}
\normalsize
Note that the quantity $V_{\bm Q}(\taubarhat)$ depends on ${\bm Q} \in {\mathbb Q}$ through the $\widetilde{M}_{ii^*}(z,z^*)$'s, which depend on ${\bm Q}$.

\begin{theorem} \label{thm2}
For any ${\bm Q} \in {\mathbb Q}$, where ${\mathbb Q}$ represents a class of psd matrices of order $N$ satisfying (\ref{eq:Q_cond}),
\begin{enumerate}
\item [(1)] The sampling variance of $\hat{\bar {\tau}}$ given by (\ref{eq:thm1varest}) of Theorem \ref{thm1} can be decomposed as:
\begin{equation}
var(\taubarhat) = V_{\bm Q}(\taubarhat) - {\bm \tau}^{\prime} {\bm Q} {\bm \tau}, \label{eq:thm2vardecomp}
\end{equation}
where ${\bm \tau} = (\tau_1, \ldots, \tau_N)^{\prime}$ is the vector of unit-level treatment contrasts given by (\ref{eq:unitlevelcontrast}), and $V_{\bm Q}(\taubarhat)$ is given by (\ref{eq:upperbound}).

\item [(2)] An upper bound of $var(\taubarhat)$ is given by $V_{\bm Q}(\taubarhat)$.

\item [(3)] This bound is attained for every treatment contrast $\bar{\tau}$ if and only if:
\begin{equation}
{\bm Q} \left \{ {\bm Y}(z) - {\bm Y}(z^*) \right\} = 0, \ \ \ z, z^* \in Z. \label{eq:thm2cond}
\end{equation}
\end{enumerate}
\end{theorem}
\noindent The proof of Theorem \ref{thm2} is in the Appendix.

\medskip

\begin{remark} \label{remark: implications of Theorem 2}

Theorem \ref{thm2} has the following implications:

\begin{enumerate}
\item[(i)] Part (1) of Theorem \ref{thm2} provides a generalization of the Neymanian decomposition of $var(\hat{\bar{\tau}})$ through the class of matrices ${\mathbb Q}$. Because the Neymanian decomposition is achieved using the matrix ${\bm Q}_{\text{strict}}$ defined by (\ref{eq:Q^*}), and ${\bm Q}_{\text{strict}} \in {\mathbb Q}$ as noted earlier, it follows that the Neymanian decomposition is a special case of the decomposition (\ref{eq:thm2vardecomp}) in part (1).

\item[(ii)] By standard arguments in finite population sampling, the component $V_{\bm Q}(\taubarhat)$ in the decomposition of $var(\hat{\bar{\tau}})$, given by (\ref{eq:upperbound}), can be unbiasedly estimated by the quantity:
\small
\begin{eqnarray}
\hat{V}_{\bm Q}(\taubarhat) &=& M + \sum_{z \in Z} \sum_{i \in T(z)} \frac{1}{\pi_i(z)} \left( M_i \left( z \right) Y_i \left( z \right) + M_{ii} \left(z \right) \left( Y_i \left( z \right) \right)^2 \right) \nonumber \\
&+& \sum_{z \in Z} \sum_{z^* \in Z} \sum_{i \in T(z) } \sum_{i^* (\ne i) \in T(z^*)} \frac{\widetilde{M}_{ii^*}(z, z^*)}{\pi_{ii^*}(z, z^*)}  Y_i(z) Y_{i^*}(z^*), \label{eq:VQ_est}
\end{eqnarray}
\normalsize
provided $\pi_i(z) > 0$, which is satisfied by assumption (\ref{eq:NScondition}), and
\begin{equation}
\pi_{ii^*}(z, z^*) > 0, \;\ \mbox{whenever} \;\ \widetilde{M}_{ii^*}(z, z^*) \ne 0. \label{eq:second-order-cond}
\end{equation}
Part (2) of Theorem \ref{thm2} thus implies that $\hat{V}_{\bm Q}(\taubarhat)$ overestimates $var(\hat{\bar{\tau}})$.

\item[(iii)] Part (3) of Theorem \ref{thm2} implies that for any ${\bm Q} \in {\mathbb Q}$ satisfying (\ref{eq:thm2cond}), $var(\hat{\bar{\tau}})$ equals $V_{\bm Q}(\taubarhat)$, and thus can be unbiasedly estimated by $\hat{V}_{\bm Q}(\taubarhat)$ given by (\ref{eq:VQ_est}), provided (\ref{eq:second-order-cond}) holds. Remark \ref{remark: Qstrictconnection} implies that condition (\ref{eq:thm2cond}), with ${\bm Q} = {\bm Q}_{\text{strict}}$, is equivalent to strict additivity. For any other ${\bm Q} \in \mathbb{Q}$, it is, therefore, immediate that condition (\ref{eq:thm2cond}) is no more demanding than strict additivity because, by (\ref{eq:Q^*}) and the first equation in (\ref{eq:Q_cond}), the rows of ${\bm Q}$ are spanned by those of ${\bm Q}_{\text{strict}}$. Indeed, for any such ${\bm Q}$, (\ref{eq:thm2cond}) is a milder condition than strict additivity whenever the row space of ${\bm Q}$ is a proper subspace of the row space of ${\bm Q}_{\text{strict}}$, i.e., $\text{rank}({\bm Q}) < N-1$. Thus, (\ref{eq:thm2cond}) of Theorem \ref{thm2} provides a condition that is milder than strict additivity but still permits unbiased estimation of $var(\hat{\bar{\tau}})$.

\end{enumerate}

\end{remark}

    The points noted in Remark \ref{remark: implications of Theorem 2} underscore the central role played by (\ref{eq:thm2cond}) and (\ref{eq:second-order-cond}) in unbiased sampling variance estimation. Condition (\ref{eq:second-order-cond}) leads to the estimator (\ref{eq:VQ_est}), which is unbiased for $V_{\bm Q}(\taubarhat)$   but has a non-negative bias for $var(\taubarhat)$. Under condition (\ref{eq:thm2cond}), (\ref{eq:VQ_est}) is unbiased for $var(\taubarhat)$ as well. A crucial difference between these conditions is that condition (\ref{eq:thm2cond}) involves the potential outcomes but not the assignment mechanism, and condition (\ref{eq:second-order-cond}) involves the assignment mechanism but not the potential outcomes. Both conditions, however, involve the matrix ${\bm Q} \in {\mathbb Q}$. Henceforth, we will refer to condition (\ref{eq:second-order-cond}) as the \emph{\textbf{second-order assignment probability condition}} (SAP condition) and condition (\ref{eq:thm2cond}) as the \emph{\textbf{generalized additivity condition}} (GA condition). We examine both of these conditions in more detail in the context of Examples 1-3 in Section \ref{s:conditions}.

\subsection{Sampling covariance and its estimation} \label{ss:estNeycov}

The ideas in Section \ref{ss:estNeyvar} also apply to the estimation of sampling covariance. Consider two treatment contrasts $\bar{\tau}_l = \sum_{z \in Z} g_l(z) \bar{Y}(z)$ and their LUEs   $\hat{\bar{\tau}}_l$, $l =1,2$. Using Lemma \ref{lemma1}, we have Theorem \ref{thm3}, which parallels Theorem \ref{thm1}:

\begin{theorem} \label{thm3}
Under a GAM, the sampling covariance between $\hat{\bar{\tau}}_1$ and $\hat{\bar{\tau}}_2$ is:
\small
\begin{eqnarray}
cov(\hat{\bar{\tau}}_1, \hat{\bar{\tau}}_2) &=& R + \sum_{z \in Z} \sum_{i=1}^N \left( R_i \left( z \right) Y_i \left( z \right) + R_{ii} \left(z \right) \left( Y_i \left( z \right) \right)^2 \right) \nonumber \\
&+& \sum_{z \in Z} \sum_{z^* \in Z} \sum_{i=1}^N \sum_{i^* (\ne i) = 1}^N R_{ii^*} (z, z^*) Y_i(z) Y_{i^*}(z^*) - \bar{\tau}_1 \bar{\tau}_2, \label{eq:thm3covest}
\end{eqnarray}
where
\begin{eqnarray}
R &=& \sum_{z \in Z} \sum_{z^* \in Z} g_1(z) g_2(z^*) A(z, z^*), \nonumber \\
R_i(z) &=& g_1(z) \sum_{z^* \in Z} g_2(z^*) A_i^{(1)} \left(z, z^* \right) + g_2(z) \sum_{z^* \in Z} g_1(z^*) A_i^{(2)} \left(z^*,z \right), \nonumber \\
R_{ii}(z) &=& g_1(z) g_2(z) B_{ii}(z,z), \ \ \ R_{ii^*}(z, z^*) = g_1(z) g_2(z^*) B_{ii^*}(z, z^*), \label{eq:R}
\end{eqnarray}
for $z, z^* \in Z$ and $i, i^*(\ne i) = 1, \ldots, N$.
\end{theorem}

\bigskip

An unbiased estimator of $cov(\hat{\bar{\tau}}_1, \hat{\bar{\tau}}_2)$ can be derived using ideas similar to those used in Section \ref{ss:estNeyvar} for unbiased estimation of the sampling variance of $\taubarhat$. For any ${\bm Q} \in \mathbb{Q}$, define, similar to (\ref{eq:tildeM}),
\begin{equation}
\widetilde{R}_{ii^*}(z, z^*) = g_1(z) g_2(z^*) \left\{ B_{ii^*} (z, z^*) + q_{ii^*} - (1/N^2) \right\}. \label{eq:tildeR}
\end{equation}
\noindent Proceed, as in the proof of Theorem \ref{thm2}, part (1), to obtain
$$ cov(\hat{\bar{\tau}}_1, \hat{\bar{\tau}}_2) = C_{\bm Q}(\hat{\bar{\tau}}_1,\hat{\bar{\tau}}_2) - {\bm \tau}^{\prime}_{[1]}{\bm Q}{\bm \tau}_{[2]}, $$
where ${\bm \tau}_{[1]}$ and ${\bm \tau}_{[2]}$ are column vectors of unit-level treatment contrasts corresponding to $\bar{\tau}_1$ and $\bar{\tau}_2$ respectively, and
\begin{eqnarray*}
C_{\bm Q}(\hat{\bar{\tau}}_1,\hat{\bar{\tau}}_2) = &R& + \sum_{z \in Z} \sum_{i=1}^N \left( R_i \left( z \right) Y_i \left( z \right) + R_{ii} \left(z \right) \left( Y_i \left( z \right) \right)^2 \right) \nonumber \\
&+& \sum_{z \in Z} \sum_{z^* \in Z} \sum_{i=1}^N \sum_{i^* (\ne i) = 1}^N \widetilde{R}_{ii^*} (z, z^*) Y_i(z) Y_{i^*}(z^*).
\end{eqnarray*}

In the spirit of the SAP condition (\ref{eq:second-order-cond}), if $\pi_{i i^*} (z, z^*)  > 0$, whenever $\widetilde{R}_{ii^*} (z, z^*) \ne  0$, then $C_{\bm Q}(\hat{\bar{\tau}}_1,\hat{\bar{\tau}}_2)$  admits an unbiased estimator along the lines of (\ref{eq:VQ_est}). If, in addition, the GA condition (\ref{eq:thm2cond}) holds, then ${\bm \tau}^{\prime}_{[1]}{\bm Q}{\bm \tau}_{[2]}$ vanishes and this estimator becomes unbiased for $cov(\hat{\bar{\tau}}_1, \hat{\bar{\tau}}_2)$ as well. Thus GA plays a crucial role in unbiased estimation of, not only the sampling variance of each treatment contrast estimator, but also the sampling covariance between two of them

\section{A closer examination of the unbiasedness conditions, with examples} \label{s:conditions}

In Section \ref{s:Neyman}, we identified two key conditions that allow unbiased estimation of the sampling variance of $\taubarhat$. The GA condition is a property of the unknown potential outcomes that cannot be known even after experimentation. In contrast, the SAP condition depends on the assignment mechanism. In this section, we explore these two conditions in more depth with respect to the examples in Section \ref{ss:examples}.

\subsection{The GA condition and choice of ${\bm Q}$} \label{ss:GA}

The GA condition, given by (\ref{eq:thm2cond}), is fulfilled if there exists a matrix ${\bm Q} \in {\mathbb Q}$, such as ${\bm Q}_{\text{strict}}$ in (\ref{eq:Q^*}), that satisfies ${\bm Q}\left( {\bm Y} \left(z \right) - {\bm Y} \left(z^* \right) \right)=0, \ z, z^* \in Z$. We now explore some other choices of ${\bm Q}$.

\begin{itemize}
\item[(a)] {\textbf{Stratum-level additivity}}: Let $N_1, \ldots, N_H \ (\ge 2)$ be integers that sum to $N$, and for $h=1, \ldots, H$, let ${\bm Q}_h$ be the square matrix of order $N_h$ having each diagonal element equal to $1/N^2$ and each off-diagonal element equal to $-1/\left(N^2\left(N_h-1 \right) \right)$. Then the psd matrix
    \begin{equation}
    {\bm Q}_{\text{strat}} = \textrm{diag}({\bm Q}_1, \ldots, {\bm Q}_H) \label{eq:Q_strata}
    \end{equation}
     satisfies (\ref{eq:Q_cond}) and therefore ${\bm Q}_{\text{strat}} \in {\mathbb Q}$. Let $\Omega_1 = \{ 1,\ldots, N_1 \}, \Omega_2 = \{N_1+1, \ldots, N_1+N_2 \}, \ldots, \Omega_H = \{\sum_{h=1}^{H-1} N_h + 1, \ldots, N \}$. For ${\bm Q}_{\text{strat}}$ defined in (\ref{eq:Q_strata}), condition (\ref{eq:thm2cond}) is equivalent to having additivity within each $\Omega_h$ which, for $H \ge 2$, is milder than strict additivity.

\item[(b)] \textbf{Whole-plot-level additivity}: Let $H \ge 2$. With $N_1, \ldots, N_H$  and $\Omega_1, \ldots, \Omega_H$ as in (a), now suppose $N$ is an integral multiple of $H$ and let $N_1= \ldots = N_H = N/H = N_0$. Take ${\bm Q}$ in partitioned form as:
    \begin{equation}
    {\bm Q}_{\text{whole-plot}} = ({\bm Q}_{hh^*})_{h,h^*=1, \ldots, H}, \label{eq:Q_whole}
    \end{equation}
where each ${\bm Q}_{hh^*}$ is a square matrix of order $N_0$, such that each element of ${\bm Q}_{hh}$  is $1/N^2$, whereas for $h \ne h^*$, each element of ${\bm Q}_{hh^*}$ is $-1/ \left( N^2 \left( H - 1 \right) \right)$. Then ${\bm Q}_{\text{whole-plot}} \in {\mathbb Q}$, and condition (\ref{eq:thm2cond}) is equivalent to the constancy of $\bar{Y}_h(z) - \bar{Y}_h(z^*)$ over $h = 1, \ldots, H$, for every $z, z^* \in Z$, where $\bar{Y}_h(z) = \sum_{i \in \Omega_h} Y_i(z)/N_0$, (i.e., the average of all potential outcomes corresponding to treatment $z$ in $\Omega_h$) for each $h$ and $z$. Hence, (\ref{eq:thm2cond}) is now equivalent to additivity of the $\bar{Y}_h(z)$, which can be viewed as additivity between the $\Omega_h$ and is again milder than strict additivity; e.g., if the $\Omega_h$ represent whole-plots as in Example 2 of Section \ref{ss:examples}, then (\ref{eq:thm2cond}) only requires between-whole-plot additivity, as defined by \cite{Zhao:2016}. Incidentally, if $N_1, \ldots, N_H$ are not all equal, then it can be impossible to find a psd counterpart of ${\bm Q}_{\text{whole-plot}}$.
\end{itemize}

Even without any split-plot type interpretation, a special case of ${\bm Q}_{\text{whole-plot}}$ given by (\ref{eq:Q_whole}) can be of interest, as will be seen later in the second example of Section \ref{ss:neat_examples}. For $H = 2$ and $N=2 N_0$, ${\bm Q}_{\text{whole-plot}}$ is:
\begin{equation}
{\bm Q}_{\text{half}} = \left(
                          \begin{array}{cc}
                            \tilde{\bm Q} & -\tilde{\bm Q} \\
                            -\tilde{\bm Q} & \tilde{\bm Q} \\
                          \end{array}
                        \right) \label{eq:Q_half}
\end{equation}
where $\tilde{\bm Q}$  is a square matrix of order $N_0$ with each element equal to $1/N^2$. In this case, the GA condition (\ref{eq:thm2cond}) is equivalent to									 \begin{equation}
\sum_{i=1}^{N_0} \left( Y_i(z) - Y_i(z^*) \right) = \sum_{i=N_0+1}^{N} \left( Y_i(z) - Y_i(z^*) \right), \ z, z^* \in Z. \label{eq:Q_half_cond}
\end{equation}
As a result, ${\bm Q}_{\text{half}}$ may be a sensible choice when the finite population can be split into two groups of equal size such that each unit in the first group has a counterpart, or match, in the second group.

\subsection{The SAP condition} \label{ss:SAP}

Using the definitions of $\widetilde{M}_{i i^*} (z, z^*)$ and $B_{i i^*} (z, z^*)$ in (\ref{eq:tildeM}) and (\ref{eq:Bij}) respectively, the SAP condition (\ref{eq:second-order-cond}), can be stated as:
\begin{equation}
\pi_{i i^*}(z, z^*) = 0 \Rightarrow g(z) g(z^*) \left\{ q_{i i^*} - (1/N^2) \right\} = 0. \label{eq:SAP2}
\end{equation}

\noindent Then a set of sufficient conditions for (\ref{eq:SAP2}) is
\begin{equation}
\pi_{i i^*}(z, z^*) = 0 \ \mbox{for at least one pair} \ z, z^* \in Z \Rightarrow q_{i i^*} = 1/N^2, \ i, i^* (\ne i) = 1, \ldots, N. \label{eq:SAP_suff}
\end{equation}
These conditions are also necessary if $g(z) \ne 0$ for every $z \in Z$. Instead of (\ref{eq:SAP2}), it will often help to invoke (\ref{eq:SAP_suff}) because it does not involve $g(z)$ and hence is relatively simpler. Of course, both (\ref{eq:SAP2}) and (\ref{eq:SAP_suff}) are trivially satisfied if the $\pi_{i i^*}(z, z^*)$ are all positive, which is the case for stratified random assignment (Example 1 in Section \ref{ss:examples}), as can be seen from equations (\ref{eq:pi_ijexample1a}) and (\ref{eq:pi_ijexample1b}) in the Appendix.

In contrast, for the split-plot assignment (Example 2) some of the $\pi_{i i^*}(z, z^*)$ vanish even when $r_1(z_1) \ge 2$ for every $z_1$, as assumed earlier. From (\ref{eq:pi_ijexample2a}) and (\ref{eq:pi_ijexample2b}), one can see that (\ref{eq:SAP_suff}) holds in this example for ${\bm Q}_{\text{whole-plot}}$ in (\ref{eq:Q_whole}) with appropriate labeling of units, but not for ${\bm Q}_{\text{strict}}$ in (\ref{eq:Q^*}) or ${\bm Q}_{\text{strat}}$ in (\ref{eq:Q_strata}).

Finally, in Example 3, by (\ref{eq:pi_ijexample3a}) and (\ref{eq:pi_ijexample3b}), for every $i, i^*(\ne i)$, either  $\pi_{i i^*}(z,z)=0$ for all $z$, or $\pi_{i i^*}(z, z^*) = 0$ for all $z, z^*(\ne z)$, so that (\ref{eq:SAP2}) forces each off-diagonal element of ${\bm Q}$ to be $1/N^2$, because any treatment contrast has $g(z) \ne 0$ for at least two distinct $z$. Because no such ${\bm Q}$ can be found in $\mathbb{Q}$, the assignment mechanism in Example 3 cannot use Theorem \ref{thm1} via (\ref{eq:VQ_est}) for unbiased variance estimation, which is anticipated because unicluster sampling designs, which motivate this mechanism, are well-known in finite population sampling to have the same deficiency.

\subsection{The strength of the generalized approach to variance estimation using the GA and SAP conditions} \label{ss:strength}

We now indicate how the findings in Sections \ref{ss:GA} and \ref{ss:SAP} reveal the strength of our approach. If the strata $\Omega_h$ in example (a) exhibit within stratum additivity, then with ${\bm Q}_{\text{strat}}$ in (\ref{eq:Q_strata}), the GA condition is met. Consequently, by Theorem \ref{thm2}, the sampling variance of any treatment contrast estimator can be unbiasedly estimated using (\ref{eq:VQ_est}), whenever the assignment mechanism satisfies the SAP condition. This is possible for any assignment mechanism, and not just the mechanism in Example 1 of Section \ref{ss:examples}, regardless of the LUEs chosen for the treatment means. Without the approach adopted here, such a general conclusion does not emerge as a na\'{\i}ve extension of the previous results in causal inference literature, e.g., those in \cite{Dasgupta:2015} that pertain to the special case of a balanced completely randomized assignment mechanism along with mean per unit estimators, even though within stratum additivity reduces to strict additivity for $H =1$. Similarly, if the $\Omega_h$ in (b) of Section \ref{ss:GA} are whole-plots and between whole-plot additivity holds, then with ${\bm Q}_{\text{whole-plot}}$ in (\ref{eq:Q_whole}), the GA condition is met. Consequently, (\ref{eq:VQ_est}) can be used for unbiased variance estimation for \emph{all} LUEs of the treatment means and under \emph{any} assignment mechanism satisfying (\ref{eq:SAP_suff}), and not just the mechanism in Example 2 of Section \ref{ss:examples}. It is this generality that makes Theorem \ref{thm2} and the resultant estimator (\ref{eq:VQ_est}) important.

The idea of imposing a version of additivity milder than strict additivity (by choosing an appropriate matrix ${\bm Q}$ that is different from ${\bm Q}_{\text{strict}}$) to estimate the sampling variance of $\taubarhat$ under less restrictive conditions appears to be attractive. However, unless there is a firm belief in the validity of the GA condition (\ref{eq:thm2cond}), the consequences of departures from it should be considered when choosing ${\bm Q}$. We discuss this aspect further in Section \ref{s:bias}.

\subsection{Some examples for specific assignment mechanisms and choice of ${\bm Q}$} \label{ss:neat_examples}

We now illustrate, how the use of HT-estimators of the treatment means along with specific choices of ${\bm Q}$ that satisfy the SAP condition lead to intuitive algebraic expressions of sampling variances of treatment contrasts and their unbiased estimators when the GA condition is met. We also indicate how some of these expressions relate to existing results.

\bigskip

\noindent \textbf{Example 1 (stratified assignment mechanism) continued}. For each $z$, consider the HT-estimator defined by (\ref{eq:HT}), which by (\ref{eq:pi_iexample1}), reduces to $\YbarhatHTz = \sum_{h=1}^H N_h \Ybarhathz/N$, where $\Ybarhathz = \sum_{i \in T_h(z)} Y_i(z) / r_h(z)$ is the average of the observed outcomes from the $r_h(z)$ units assigned to treatment $z$ in stratum $\Omega_h$, and $T_h(z)$ is the set of these $r_h(z)$ units. From standard results in finite population sampling theory, for the HT-estimator,
\begin{eqnarray}
A(z, z^*) &=& A_i^{(1)}(z, z^*) = A_i^{(2)}(z, z^*) = 0, \nonumber \\
B_{ii}(z, z) &=& 1/ \left( N^2 \pi_i(z) \right), \;\ B_{ii^*}(z, z^*) = \pi_{ii^*}(z, z^*)/ \left( N^2 \pi_i \left(z \right) \pi_{i^*} \left(z^* \right) \right). \label{eq:HT_AB}
\end{eqnarray}
Substituting the first and second-order assignment probabilities for Example 1 (from equations (\ref{eq:pi_iexample1}) to (\ref{eq:pi_ijexample1b}) in the Appendix) into (\ref{eq:HT_AB}), gives, for any $i \in \Omega_h, \ i^* (\ne i) \in \Omega_{h^*}$,
\begin{eqnarray}
B_{ii}(z, z) &=& \frac{N_h}{N^2 r_h(z)}, \ \ B_{ii^*}(z,z) = \frac{N_h \left( r_h(z) - 1\right)}{N^2 (N_h-1) r_h(z)}, \ \mbox{if} \ h=h^* \nonumber  \\
B_{ii^*}(z, z^*) &=& \frac{N_h}{N^2(N_h-1)} \ \mbox{if} \ h=h^*, z \ne z^*, \ \ B_{ii^*}(z, z^*) = \frac{1}{N^2} \ \mbox{if} \ h \ne h^* .
\label{eq:HT_strat_AB}
\end{eqnarray}
Now, label the units so that $\Omega_1 = \{1, \ldots, N_1\}$, $\Omega_2 = \{N_1+1, \ldots, N_1+N_2\}$, etc. Then with ${\bm Q} = {\bm Q}_{\text{strat}}$ as in (\ref{eq:Q_strata}), substitution of (\ref{eq:HT_strat_AB}) into (\ref{eq:tildeM}) yields
\[   \widetilde{M}_{ii^*}(z, z^*) = \left \{ \begin{array}{cc}
                                -\frac{\left( g(z) \right)^2 N_h}{N^2 (N_h-1) r_h(z)},  & \ \mbox{if} \ z=z^* \ \mbox{and} \ i,i^* \in \Omega_h \ \mbox{for some} \ h  \\
                                0 & \mbox{otherwise.}
                              \end{array} \right.
\]
Similar expressions for $M, M_i(z), M_{ii}(z)$ can be obtained by substituting (\ref{eq:HT_strat_AB}) into (\ref{eq:M}). Finally, substituting the expressions of $M, M_i(z), M_{ii}(z)$ and $\widetilde{M}_{ii^*}(z, z^*)$ into (\ref{eq:upperbound}) and (\ref{eq:VQ_est}), and after some simplification, we obtain:
\begin{eqnarray}
V_{{\bm Q}_{\text{strat}}}(\taubarhatHT) &=& \frac{1}{N^2} \sum_{z \in Z} \sum_{h=1}^H  \left( g\left(z \right) \right)^2 \frac{ N_h^2 }{ r_h(z) } S_h(z,z), \label{eq:VQexample1} \\
\hat{V}_{{\bm Q}_{\text{strat}}}(\taubarhatHT) &=& \frac{1}{N^2} \sum_{z \in Z} \sum_{h=1}^H  \left( g\left(z \right) \right)^2 \frac{ N_h^2 }{ r_h(z) } \hat{S}_h(z,z), \label{eq:VQhatexample1}
\end{eqnarray}
where $ S_h(z,z) = \sum_{i \in \Omega_h} \left( Y_i \left(z \right) - \bar{Y}_h \left(z \right) \right)^2 / (N_h-1)$ is the variance (with divisor $N_h-1$) and $\bar{Y}_h(z) = \sum_{ i \in \Omega_h} Y_i(z) / N_h$ is the average of all potential outcomes of units in stratum $\Omega_h$ for treatment $z$, and $$ \hat{S}_h(z,z) = \sum_{i \in T_h(z)} \left( Y_i \left(z \right) - \Ybarhathz \right)^2 / (r_h(z)-1) $$ is an unbiased estimator of $S_h(z,z)$, calculated as the sample variance of observed outcomes in stratum $\Omega_h$ that correspond to treatment $z$.

Also ${\bm \tau}^{\prime} {\bm Q}_{\text{strat}} {\bm \tau} = \sum_{h=1}^H N_h S_h(\tau, \tau)/N^2$, where
$S_h(\tau, \tau) = \sum_{i \in \Omega_h} \left( \tau_i - \taubar^{(h)} \right)^2 / (N_h - 1)$ is the variance (with divisor $N_h-1$) and $\taubar^{(h)} = \sum_{i \in \Omega_h} \tau_i / N_h$ is the average of the unit-level treatment contrasts within stratun $\Omega_h$. Consequently, by Theorem \ref{thm2}, part (1), we obtain
\begin{equation}
var(\taubarhatHT) = \frac{1}{N^2} \sum_{h=1}^H N_h \left( \sum_{z \in Z} \left( g(z) \right)^2 \frac{N_h}{r_h(z)} S_h(z,z) - S_h(\tau, \tau)  \right), \label{eq:vartauhatexample1}
\end{equation}

When $H=1$, a stratified assignment mechanism reduces to a completely randomized assignment, and ${\bm Q}_{\text{strat}}$ reduces to ${\bm Q}_{\text{strict}}$. Consequently, expressions (\ref{eq:VQexample1}), (\ref{eq:VQhatexample1}) and (\ref{eq:vartauhatexample1}) reduce to the expressions derived by \cite{Dasgupta:2015}. When $Z = \{0,1\}$, (\ref{eq:vartauhatexample1}) further reduces to the Neymanian decomposition (\ref{eq:Ney_var}) discussed in Section \ref{ss:LUE}.

\bigskip

\noindent \textbf{Example on completely randomized assignment using ${\bm Q}_{\text{half}}$}. Let $N = 2 N_0$ so that ${\bm Q}_{\text{half}}$ in (\ref{eq:Q_half}) is well-defined. Consider a completely randomized assignment that allocates $r(z) \ (\ge 2), z \in Z,$ units to treatment $z$, where the $r(z)$'s sum to $N$. As in the previous example, for each $z$, consider the HT-estimator $\YbarhatHTz = \sum_{i \in T(z)} Y_i(z) / r(z)$. Because the assignment mechanism in this example is a special case of the previous example with $H=1$, dropping the index $h$ from the expressions in (\ref{eq:HT_strat_AB}) results in
\begin{eqnarray*}
B_{ii}(z, z) &=& \frac{1}{N r(z)}, \ \ B_{ii^*}(z,z) = \frac{ r(z) - 1 }{N (N-1) r(z)},  \\
B_{ii^*}(z, z^*) &=& \frac{1}{N(N-1)} \ \mbox{if} \ z \ne z^*.
\label{eq:HT_half_AB}
\end{eqnarray*}
Substituting the above expressions into (\ref{eq:M}) and (\ref{eq:tildeM}), and further substituting the resultant expressions of $M$'s and $\widetilde{M}$'s into (\ref{eq:upperbound}) and (\ref{eq:VQ_est}), after considerable algebra, we obtain
\small
\begin{eqnarray*}
V_{{\bm Q}_{\text{half}}}(\taubarhatHT) &=& \sum_{z \in Z} \frac{ \left( g\left(z \right) \right)^2}{r(z)} S(z,z) + \frac{1}{N(N-1)} \sum_{i=1}^N \sum_{i^* (\ne i)=1}^N \tau_i \tau_{i^*} - \taubar^{(1)} \taubar^{(2)}, \\
\hat{V}_{{\bm Q}_{\text{half}}}(\taubarhatHT) &=& \taubarhat^2 - \frac{4(N-1)}{N} \sum_{z \in Z} \frac{ \left( g\left(z \right) \right)^2}{r(z)\left( r(z)-1 \right)} L_1(z) L_2(z) \\
&-& \frac{4(N-1)}{N} \sum_{z \in Z} \sum_{z^* (\ne z) \in Z} \frac{ g(z) g(z^*) }{r(z) r(z^*)} L_1(z) L_2(z^*),
\end{eqnarray*}
\normalsize
where
$$ S(z,z) = \sum_{i=1}^N \left( Y_i(z) - \bar{Y}(z) \right)^2/(N-1)$$ is the variance (with divisor $N-1$) of potential outcomes of all units for treatment $z$, $$ \taubar^{(1)} =  \frac{1}{N_0}  \sum_{i=1}^{N_0} \tau_i  \ \  \mbox{and} \ \ \taubar^{(2)} = \frac{1}{N_0}  \sum_{i=N_0+1}^{N} \tau_i  $$ are the averages of unit-level estimands of units $\{1, \ldots, N_0 \}$ and $\{N_0+1, \ldots, N \}$ respectively, and $L_1(z)$ and $L_2(z)$ are the sums of observed outcomes of units, respectively in $\{1, \ldots, N_0\}$ and $\{N_0+1, \ldots, N\}$, that are exposed to treatment $z$.

The estimator $\hat{V}_{{\bm Q}_{\text{half}}}(\taubarhatHT)$ unbiasedly estimates $var(\taubarhatHT)$ if and only if (\ref{eq:Q_half_cond}) holds. Also, note that ${\bm \tau}^{\prime} {\bm Q}_{\text{half}} {\bm \tau} = \left( \taubar^{(1)} - \taubar^{(2)} \right)^2/4$. Therefore, by Theorem \ref{thm2}, part (1),
\begin{eqnarray*}
var(\taubarhatHT) = V_{{\bm Q}_{\text{half}}}(\taubarhatHT) - {\bm \tau}^{\prime} {\bm Q}_{\text{half}} {\bm \tau} = \sum_{z \in Z} \frac{ \left( g\left(z \right) \right)^2}{r(z)} S(z,z) - \frac{1}{N(N-1)} \sum_{i=1}^N (\tau_i - \taubar)^2,
\end{eqnarray*}
which is again in agreement with the results derived in \cite{Dasgupta:2015}, who decomposed $var(\taubarhatHT)$ using ${\bm Q}_{\text{strict}}$ rather than ${\bm Q}_{\text{half}}$.

\section{More on additivity: Is the Neymanian strategy ``optimal''?} \label{s:bias}

With a known assignment mechanism, if there are strong reasons to believe that there exists a psd ${\bm Q} \in \mathbb{Q}$ satisfying the SAP and GA conditions, then the results in Section \ref{ss:estNeyvar} suggest that one can use the $\hat{V}_{\bm Q}(\taubarhat)$, obtained from (\ref{eq:tildeM}) and (\ref{eq:VQ_est}), to unbiasedly estimate $var({\taubarhat})$ for any treatment contrast $\taubar$. For instance, with the stratified assignment mechanism in Example 1, if one is confident about within stratum additivity, then as discussed in (a) of Section \ref{ss:GA}, ${\bm Q}_{\text{strat}}$ in (\ref{eq:Q_strata}) can be used for unbiased variance estimation.

The GA condition (\ref{eq:thm2cond}), however, involves the potential outcomes, at least half of which are not observed and, as a result, there can be situations where one is not sure about the validity of such condition for any ${\bm Q}$. Then, in order to choose ${\bm Q}$, one should consider the consequences of departures from (\ref{eq:thm2cond}) in terms of the ensuing bias in variance estimation, with a view to limiting it.

\subsection{Positive second-order assignment probabilities} \label{ss:bias_allpositive}

We begin with assignment mechanisms for which the second-order assignment probabilities are all positive, a property in the survey literature that describes a measurable survey design \citep{Kish:1965}. A simplifying feature then is that the SAP condition is irrelevant, because it is trivially met by every ${\bm Q} \in \mathbb{Q}$. Consider such a psd matrix ${\bm Q}$. By Theorem \ref{thm2}, part (1), use of the resulting $\hat{V}_{\bm Q}(\taubarhat)$ as an estimator of $var(\taubarhat)$ will lead to a non-negative bias of magnitude ${\bm \tau}^{\prime} {\bm Q} {\bm \tau}$ in the absence of any knowledge about the GA condition (\ref{eq:thm2cond}). To understand the implications of choosing such a ${\bm Q}$ and the resultant estimator $\hat{V}_{\bm Q}(\taubarhat)$, consider three scenarios: (i) strict additivity holds (which implies that all milder versions of additivity also hold), (ii) strict additivity does not hold, but the GA condition (\ref{eq:thm2cond}) holds for some ${\bm Q}^* \in \mathbb{Q}$ (e.g., ${\bm Q}_{\text{strat}}$) (iii) Neither ${\bm Q}_{\text{strict}}$ nor ${\bm Q}^* \in \mathbb{Q}$ satisfies the GA condition (\ref{eq:thm2cond}). We consider two possible estimators of $var(\taubarhat)$: First, the generalized Neymanian variance estimator $\hat{V}_{{\bm Q}_{\text{strict}}}(\taubarhat)$, and second, $\hat{V}_{{\bm Q}^*}(\taubarhat)$ based on ${\bm Q}^*$. Table \ref{tab:bias} shows the bias of estimation associated with these situations and two choices for estimators.

\begin{table}[htbp]
\centering \small
\caption{Bias in the estimation of $var(\taubarhat)$ under six different scenarios} \label{tab:bias}
\begin{tabular}{c|cc}
\hline
Truth  & \multicolumn{2}{|c}{Variance Estimator} \\
       &  $\hat{V}_{{\bm Q}_{\text{strict}}}(\taubarhat)$ (Neyman) & $\hat{V}_{{\bm Q}^*}(\taubarhat)$ \\ \hline
${\bm Q}_{\text{strict}}$ satisfies (\ref{eq:thm2cond})  (strict additivity) &  0 & 0  \\
${\bm Q}^*$ satisfies (\ref{eq:thm2cond}) but not ${\bm Q}_{\text{strict}}$ & ${\bm \tau}^{\prime} {\bm Q}_{\text{strict}} {\bm \tau}$ & 0   \\
Neither ${\bm Q}_{\text{strict}}$ nor ${\bm Q}^*$ satisfies (\ref{eq:thm2cond}) & ${\bm \tau}^{\prime} {\bm Q}_{\text{strict}} {\bm \tau}$ & ${\bm \tau}^{\prime} {\bm Q}^* {\bm \tau}$ \\ \hline
\end{tabular}
\end{table}

Table \ref{tab:bias} shows that if strict additivity holds, then both estimators are unbiased. If milder additivity holds, but not strict additivity, then the Neymanian estimator has positive bias whereas the one based on ${\bm Q}^*$ is unbiased. However, if even milder additivity does not hold, then both sampling variance estimators are biased, and none of them has a larger bias than the other uniformly in ${\bm \tau}$. This is because, ${\bm Q}_{\text{strict}}$ and ${\bm Q}^*$ have all diagonal elements $1/N^2$, and hence unless they are equal, their difference is neither positive semidefinite nor negative semidefinite. Thus the issue of bias control becomes nontrivial, and appropriate choice of ${\bm Q} \in \mathbb{Q}$ from this perspective becomes important.

Consider minimizing the maximum possible bias to protect against the worst possible scenario, which arises when the bias ${\bm \tau}^{\prime} {\bm Q} {\bm \tau}$ is maximum with respect to ${\bm \tau}$. Whereas for any ${\bm Q}$, the set of values of ${\bm \tau}^{\prime} {\bm Q} {\bm \tau}$ over all possible ${\bm \tau}$ is not bounded above, it is also true that ${\bm \tau}$ essentially represents the same vector of unit-level treatment contrasts even after multiplication by any nonzero constant. Hence, in the spirit of E-optimality in experimental design \citep{Ehrenfeld1955}, consider choosing ${\bm Q} \in \mathbb{Q}$ to minimize
\begin{equation}
\max_{{\bm \tau}: {\bm \tau}^{\prime} {\bm \tau} = 1} {\bm \tau}^{\prime} {\bm Q} {\bm \tau} = \lambda_{\max} ({\bm Q}), \label{eq:E-opt}
\end{equation}				
where $\lambda_{\max}({\bm Q})$ is the largest eigenvalue of ${\bm Q}$. The following proposition (proof in Appendix) provides a guideline for choosing the optimal ${\bm Q}$ with respect to criterion (\ref{eq:E-opt})

\begin{proposition} \label{prop1}
For any psd matrix ${\bm Q} \in \mathbb{Q}$,  $$ \lambda_{\max} ({\bm Q}) \ge \frac{1}{N(N-1)}, $$ with equality if and only if ${\bm Q} = {\bm Q}_{\emph{strict}}$.
\end{proposition}

For ${\bm Q} = {\bm Q}_{\text{strict}}$, recall that the GA condition reduces to the strict additivity condition. Thus, if the second-order assignment probabilities are all positive, the choice of ${\bm Q}$ that makes the GA condition most demanding is also most protective against the worst scenario regarding bias.

\medskip

\begin{remark} \label{remark:Neyman_justification}
Proposition \ref{prop1} provides a justification for using the Neymanian estimator of $var(\taubarhat)$: it minimizes the maximum bias of sampling variance estimation \emph{for any assignment mechanism} for which all second-order assignment probabilities are positive.
\end{remark}

\medskip

\subsection{Case of split-plot assignment mechanism} \label{ss:bias_splitplot}

Assignment mechnisms with some vanishing second-order assignment probabilities rely on the SAP condition, which may impose messy patterns on ${\bm Q}$. Indeed, as with the assignment mechanism in Example 3 of Section \ref{ss:examples}, there may not even exist any ${\bm Q} \in \mathbb{Q}$ satisfying the SAP condition. The split-plot assignment mechanism in Example 2, however, allows an extension of the ideas in Section \ref{ss:bias_allpositive}.

\begin{proposition} \label{prop2}
Let the $N = H N_0$ units be labeled such that the $H$ whole-plots are taken as $\Omega_1 = \{1, \ldots, N_0 \}, \ldots, \Omega_H = \{(H-1) N_0 + 1, \ldots, H N_0 \}$.
\begin{itemize}
\item[(a)] For the split-plot assignment mechanism, a psd matrix ${\bm Q} \in \mathbb{Q}$ satisfies the SAP condition (\ref{eq:SAP_suff}) if and only if it is of the form ${\bm Q} = {\bm Q}_1 \bigotimes ({\bm 1}_0 {\bm 1}_0^{\prime})$, where $\bigotimes$ denotes Kronecker product, ${\bm Q}_1$ is a psd matrix of order $H$ having each row sum zero and each diagonal element $1/N^2$, and ${\bm 1}_0$  is the $N_0 \times 1$ vector of all ones.
\item[(b)] For any psd matrix ${\bm Q}$ as in (a) above, $$ \lambda_{\max} ({\bm Q}) \ge \frac{1}{N(H-1)}, $$ with equality if and only if ${\bm Q}_1$ has each diagonal element $1/N^2$ and each off-diagonal element $-1/\left(N^2(H-1)\right)$.
\end{itemize}
\end{proposition}

The proof of Proposition \ref{prop2} is in the Appendix. For the split-plot assignment mechanism, one needs to consider ${\bm Q}$ as in Proposition 2(a). With ${\bm Q}_1$ as in Proposition 2(b), such a ${\bm Q}$ equals ${\bm Q}_{\text{whole-plot}}$, the one shown in (\ref{eq:Q_whole}). Hence, ${\bm Q}_{\text{whole-plot}}$ is uniquely the best choice of ${\bm Q} \in \mathbb{Q}$, for minimizing the maximum possible bias given by (\ref{eq:E-opt}).

\section{Treatments with a factorial structure} \label{s:factorial}

The ideas discussed apply to situations where the treatments have a factorial structure. To see this, consider an $s_1 \times \ldots \times s_K$  factorial experiment with $K$ factors $F_k$, $k = 1, \ldots, K$, with the levels of $F_k$ denoted as $0, 1, \ldots, s_{k-1}$.  As earlier, the treatment combinations are $z \in Z$, and let $\bar{Y}(z) = \sum_{i=1}^N Y_i(z)/N$, where $Y_i(z)$ is the potential outcome of unit $i$   when exposed to treatment combination $z$. Let $\bar{\bm Y}$ denote the column vector of order $|Z|$ having elements $\bar{Y}(z)$, arranged lexicographically, e.g., in a $2 \times 3$ factorial experiment,
\begin{eqnarray*}
\bar{\bm Y} = \left( \bar{Y}(00), \bar{Y}(01), \bar{Y}(02), \bar{Y}(10), \bar{Y}(11), \bar{Y}(12) \right)^{\prime}.
\end{eqnarray*}
Then a typical treatment contrast belonging to a factorial effect $F_1^{x_1} \ldots F_K^{x_K}$, where $x_1, \ldots, x_K$ are binary and at least one is non-zero, is of the form ${\bm g}^{\prime} \bar{\bm Y}$, with ${\bm g}$ having the structure ${\bm g} = {\bm g}_1 \bigotimes \ldots \bigotimes {\bm g}_K$, such that each ${\bm g}_k$ is a nonnull vector having all elements equal if $x_k =0$, and elements summing to zero if $x_k =1$, $ (k = 1, \ldots, K)$ (see, for example, Chapter 2 of \cite{GM1989}). A contrast of this kind is of the form (\ref{eq:poplevelcontrast}), with $g(z)$ representing the element of ${\bm g}$ that corresponds to treatment combination $z$. Thus, all our results are applicable to treatment contrasts representing factorial main effects or interactions even under a general factorial structure for the treatments.

\section{Simulations and numerical examples} \label{s:simulations}

Some simulations demonstrate the usefulness of the proposed approach and the effects of departures from strict and mild additivity on biases of proposed sampling variance estimators. Consider a hypothetical randomized experiment with $N = 50$ units, designed to estimate the causal effects of three treatments labeled $1,2$ and 3. Assume that interest lies in the contrast $\bar{\tau} = \bar{Y}(1) - 2 \bar{Y}(2) + \bar{Y}(3)$, where $\bar{Y}(z)$ denotes the mean of the 50 potential outcomes for treatment $z$, $z \in \{1,2,3\}$. Such a contrast is often of practical interest: for qualitative treatments it can be interpreted as the causal effect of treatment 2 compared to the average effect of treatments 1 and 3, and for quantitative treatments (e.g., three levels of temperature) it represents the quadratic effect of the treatments on the outcome.

We also assume that a blocking factor (e.g., sex) is used to stratify the population of 50 units into two strata $\Omega_1$ and $\Omega_2$ of sizes 30 and 20, respectively. Accordingly, we define a finite population as a $50 \times 3$ matrix of potential outcomes, whose $i$th row is denoted by ${\bm Y}_i = \left( Y_i(1), Y_i(2), Y_i(3) \right)$, $i = 1, \ldots, 50$, where each ${\bm Y}_i$ is generated in our simulation using the trivariate normal model:
$$ {\bm Y}_i \sim N_3 \left( {\bm \mu}_h, {\bm \Sigma}_h \right), \ \ i \in \Omega_h, \ \ h=1,2, $$
where \[ {\bm \Sigma}_h = \sigma^2_h \left(
\begin{array}{ccc}
1 & \rho_h & \rho_h \\
\rho_h & 1 & \rho_h \\
\rho_h & \rho_h & 1 \\
\end{array}
\right), h=1,2.
\]
The choices of ${\bm \mu}_h, \sigma^2_h$ and $\rho_h$ for $h = 1,2$ determine the level of additivity in populations generated from the above model. We study the biases of the estimators of $var(\taubarhat)$ for six such generating models with different levels of additivity. These models are shown in Table \ref{tab:populations}. Populations generated from models I to VI are in increasing order of departure from strict additivity. Note that strict additivity holds for model I and stratum-level additivity holds for model II. Neither form of additivity holds for models III-VI; specifically, for model V, potential outcomes for the three treatments are generated from independent populations, and for model VI, potential outcomes for each pair of treatments have a negative correlation of -0.5.

\begin{table}[htb]
\centering \small
\caption{Parameters of models used to generate simulated potential outcomes} \label{tab:populations}
\begin{tabular} {c|ccccc}
\hline
Model & \multicolumn{5}{|c}{Generating parameters} \\
       & ${\bm \mu}$'s &  $\sigma^2_1$ & $\sigma^2_2$ & $\rho_1$ & $\rho_2$   \\ \hline
    I   &   ${\bm \mu}_1 = {\bm \mu}_2 = (8,7,10)$ & 2 & 2 & 1 & 1   \\
    II   &   ${\bm \mu}_1 = (10,12,14)$, ${\bm \mu}_2 = (8,6,10)$ & 2 & 3 & 1 & 1  \\
    III  &   ${\bm \mu}_1 = (10,12,14)$, ${\bm \mu}_2 = (8,6,10)$ & 2 & 3 & .2 & .9  \\
    IV   &   ${\bm \mu}_1 = (10,12,14)$, ${\bm \mu}_2 = (8,6,10)$ & 2 & 3 & .5 & .5  \\
    V   &   ${\bm \mu}_1 = (10,12,14)$, ${\bm \mu}_2 = (8,6,10)$ & 2 & 3 &  0 &  0  \\
    VI  &   ${\bm \mu}_1 = {\bm \mu}_2 = (8,7,10)$               & 3 & 3 & -.5 & -.5  \\ \hline
\end{tabular}
\end{table}

For each population, we consider a stratified assignment mechanism, which, as noted earlier, meets the SAP condition for every choice of ${\bm Q}$, and compare two estimators of $var(\taubarhat)$ obtained by substituting ${\bm Q}_{\text{strict}}$ from (\ref{eq:Q^*}) and ${\bm Q}_{\text{strat}}$ from (\ref{eq:Q_strata}) for ${\bm Q}$ in  $\hat{V}_{\bm Q}(\taubarhat)$ given by (\ref{eq:VQ_est}). By Theorem \ref{thm2}, these sampling variance estimators have biases ${\bm \tau}^{\prime} {\bm Q}_{\text{strict}} \tau$ and ${\bm \tau}^{\prime} {\bm Q}_{\text{strat}} {\bm \tau}$, respectively. We generate 100 populations from each of the models I--VI, and for each population (and hence each vector ${\bm \tau}$ of unit-level estimands) generated from a specific model, compute the biases ${\bm \tau}^{\prime} {\bm Q}_{\text{strict}} {\bm \tau}$ and ${\bm \tau}^{\prime} {\bm Q}_{\text{strat}} {\bm \tau}$, along with their ratio ${\bm \tau}^{\prime} {\bm Q}_{\text{strat}} {\bm \tau}/ {\bm \tau}^{\prime} {\bm Q}_{\text{strict}} {\bm \tau}$.

\begin{figure}[htb]
\caption{Boxplots of bias with ${\bm Q}_{\text{strict}}$ and ${\bm Q}_{\text{strat}}$ for populations generated from models III-VI} \label{fig:performance}
\centering
\begin{tabular}{cccc}
      \multicolumn{2}{l}{\includegraphics[scale=.46]{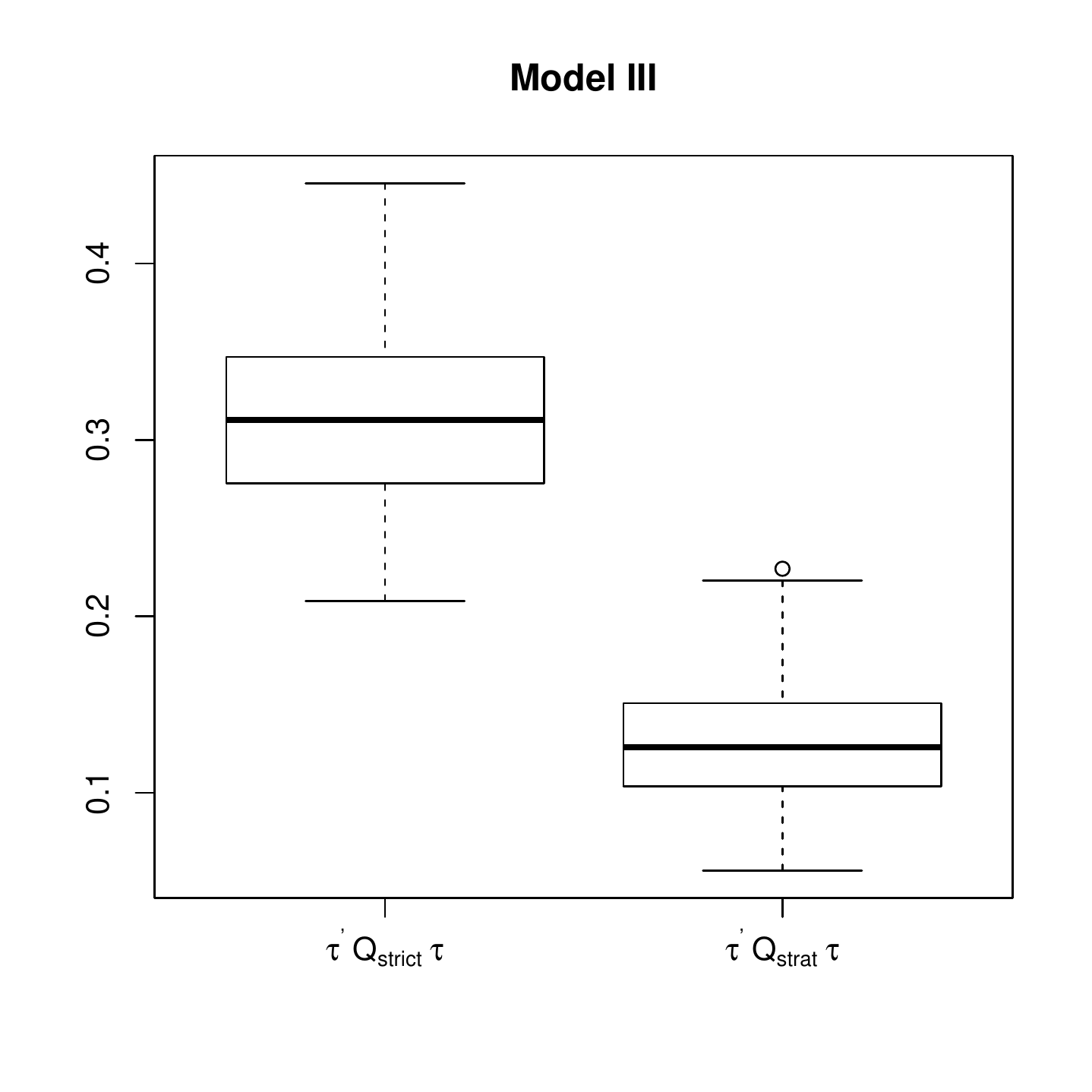}} & \multicolumn{2}{l}{\includegraphics[scale=.46]{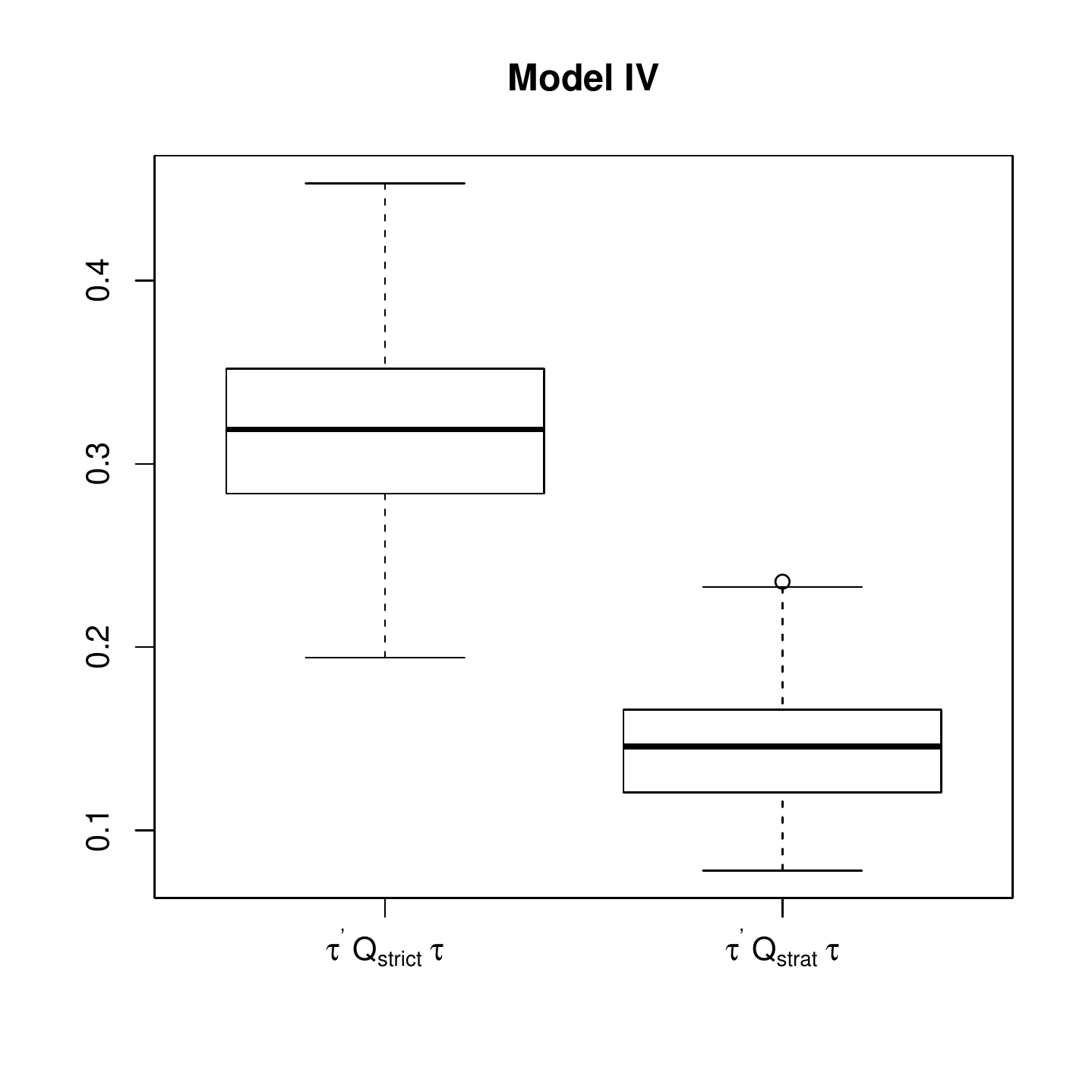}}  \\
   \multicolumn{2}{l}{\includegraphics[scale=.46]{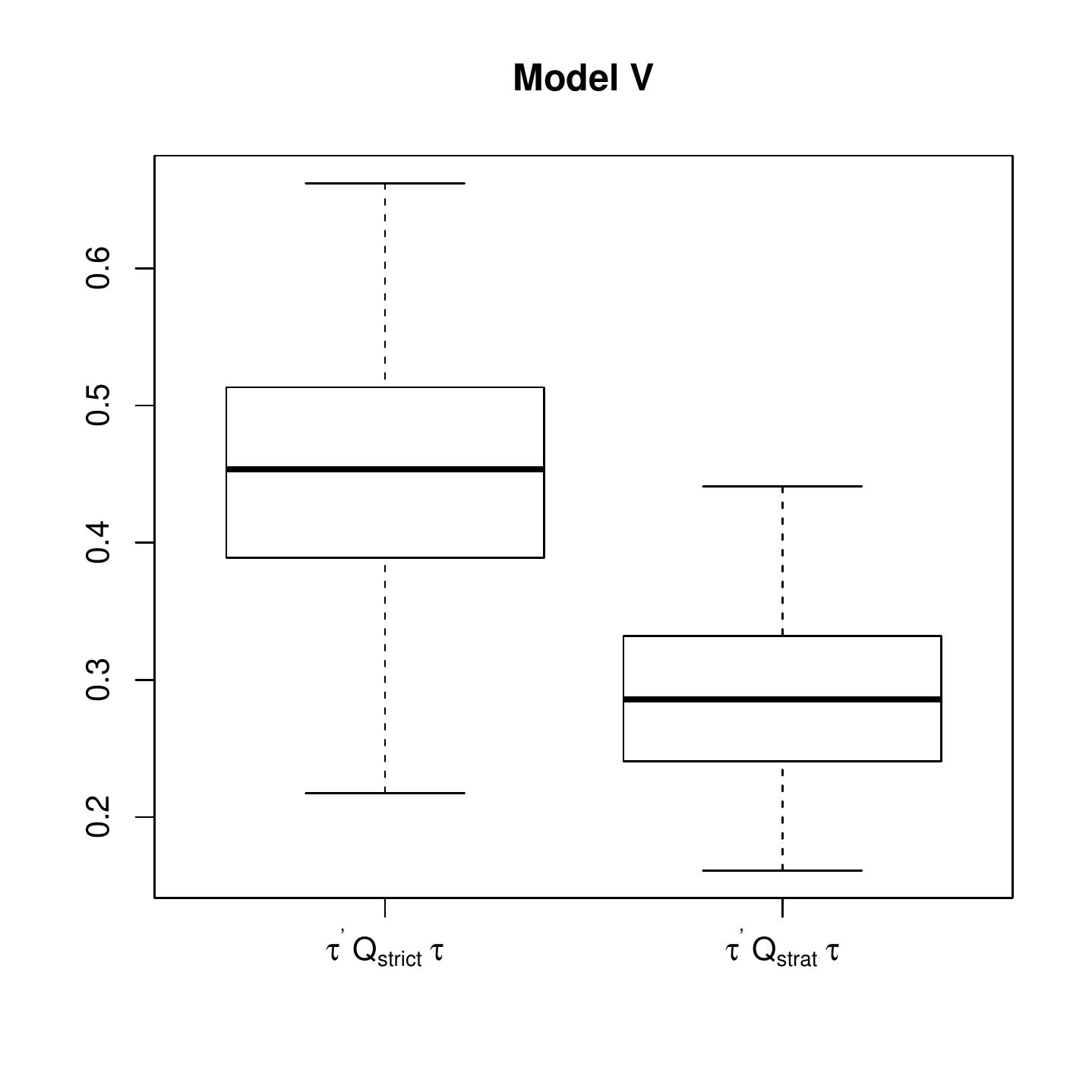}} & \multicolumn{2}{l}{\includegraphics[scale=.46]{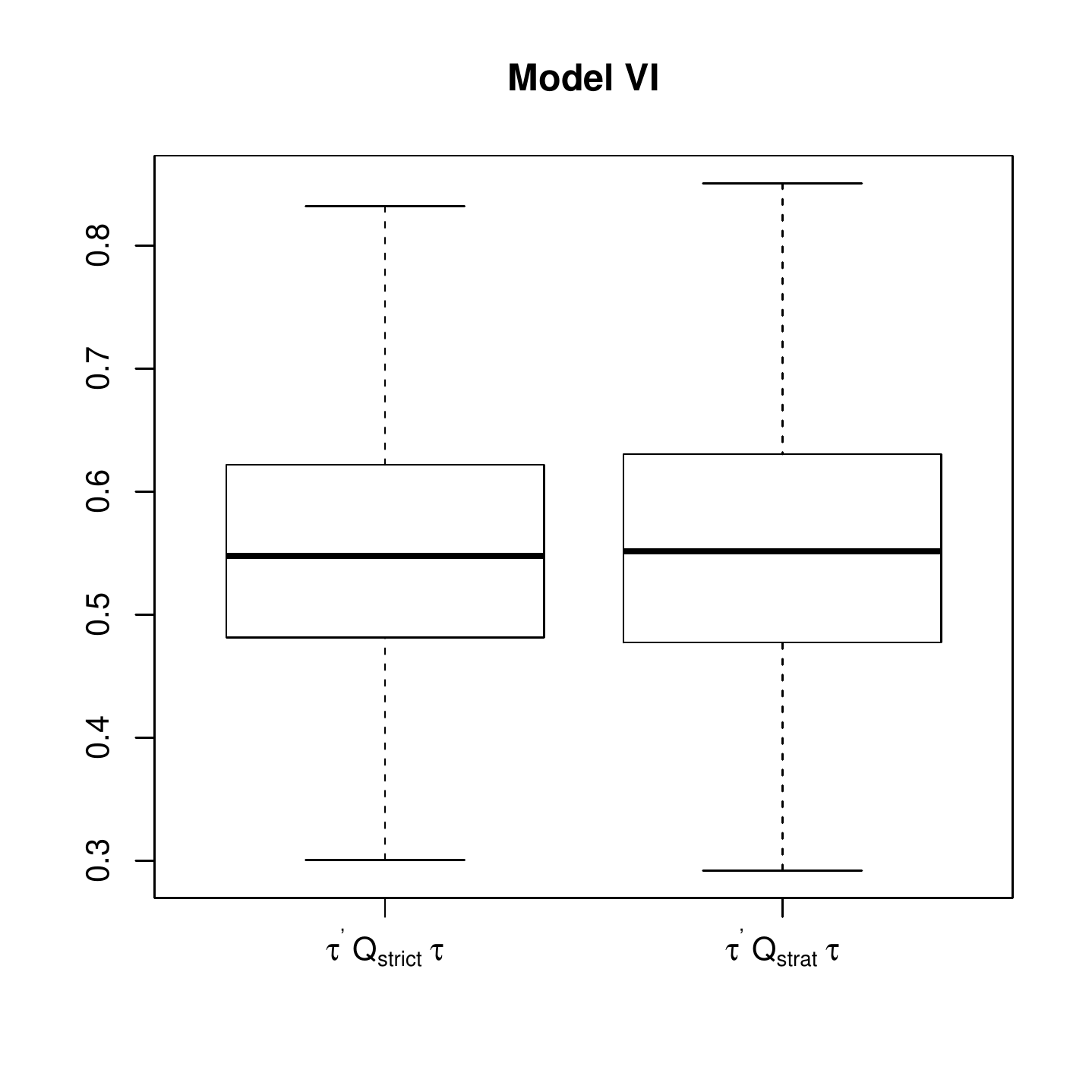}}
\end{tabular}
\end{figure}

The results are seen to follow a pattern consistent with our results. In populations generated from model I, where strict additivity holds, both sampling variance estimators lead to zero bias for each ${\bm \tau}$. For model II, where within stratum additivity holds but not strict additivity, ${\bm \tau}^{\prime} {\bm Q}_{\text{strat}} {\bm \tau}$ vanishes for every ${\bm \tau}$, but using ${\bm Q}_{\text{strict}}$ results in a small positive median bias of 0.176. The summaries (boxplots) of biases for populations generated from models III, IV, V and VI, for which neither form of additivity holds, are shown in Figure \ref{fig:performance}. The medians of the ratios of the biases are
\begin{equation}
\underset{\bm \tau} {\text{median}} \left( \frac { {\bm \tau}^{\prime} {\bm Q}_{\text{strat}} {\bm \tau}} {  {\bm \tau}^{\prime} {\bm Q}_{\text{strict}} {\bm \tau} } \right) = 0.42, 0.46, 0.63 \ \mbox{and} \ 1.01, \label{eq:bias_compare}
\end{equation}
for models III, IV, V and VI respectively. By (\ref{eq:bias_compare}), the biases arising from use of ${\bm Q}_{\text{strat}}$ are substantially smaller than those arising from use of ${\bm Q}_{\text{strict}}$, except in model VI, which entails negative correlation among pairwise columns of potential outcomes, and the largest bias among all cases, partially justifying ${\bm Q}_{\text{strict}}$ as a minimax solution. Because negatively correlated potential outcomes are unlikely to arise in practice, choosing ${\bm Q}_{\text{strict}}$ as a minimax solution can, however, be a bit pessimistic in the context of this example. A criterion of average bias, as indicated in Section 9 is an alternative choice.

\section{Discussion} \label{s:discussion}

 This article extends the existing theory of causal inference for treatment contrasts from randomized experiments in several ways and sheds light on the existing results from different perspectives. First, it develops a Neymanian inferential procedure for a \emph{general assignment mechanism} that includes, as special cases, most of the commonly used randomization and restricted randomization procedures for assigning experimental units to treatments. Second, it integrates statistical inference for causal effects with statistical inference for finite population estimands. Third, it considers multiple treatments allowing the possibility of symmetric or asymmetric factorial structure. Fourth, it proposes conditions that are milder than the well-known strict additivity condition for unbiased estimation of the sampling variance of any treatment contrast estimator. Finally, a new justification for using the Neymanian conservative sampling variance estimator is obtained. Two aspects that may be of additional interest, but involve a \emph{superpopulation perspective} rather than the finite population approach adopted in this paper, are briefly indicated below.

The first of these concerns a criterion of average or expected bias, rather than that of maximum possible bias considered in Section \ref{s:bias}, the expectation being with respect to a prior distribution on ${\bm \tau}$, as induced by a superpopulation specification on the potential outcomes. This calls for choosing ${\bm Q} \in \mathbb{Q}$ so as to minimize $E({\bm \tau}^{\prime} {\bm Q} {\bm \tau}) = tr({\bm Q} {\bm \Gamma})$, where ${\bm \Gamma} = E({\bm \tau}{\bm \tau}^{\prime})$ and the expectation is defined with respect to the superpopulation model. Although this optimization problem does not seem to admit an analytical solution in general, preliminary studies suggest that a semidefinite programming formulation should help. In contrast to Proposition \ref{prop1}, even if the second-order assignment probabilities are all positive, the resulting optimal ${\bm Q}$, now dependent on ${\bm \Gamma}$, can be quite different from ${\bm Q}_{\text{strict}}$.

From the perspective of a superpopulation model, it is also of interest to explore an optimal inference strategy, i.e., an optimal combination of an assignment mechanism and LUEs $\Ybarhatz, \ z \in Z$, that minimizes the model-expected total sampling variance of estimators of treatment contrasts of interest. Our general framework would allow borrowing ideas from finite population sampling for this purpose.

Work is currently in progress on the problems mentioned above and will be reported elsewhere.

\section*{Appendix}

\noindent \textbf{Assignment probabilities in Example 1:}

\medskip

\noindent For $i \in \Omega_h$, $h = 1, \ldots, H$, and $z \in Z$,
\begin{equation}
\pi_i(z) = r_h(z)/N_h. \label{eq:pi_iexample1}
\end{equation}

\noindent For $i, i^*(\ne i) \in \Omega_h$, $h = 1, \ldots, H$, and $z, z^* \in Z$,
\begin{eqnarray}
\pi_{ii^*}(z, z) &=& \frac{r_h(z) \left( r_h(z) - 1 \right) }{N_h (N_h-1)}, \nonumber \\
\pi_{ii^*}(z, z^*) &=&  \frac{r_h(z) r_h(z^*)}{N_h (N_h-1)}, \ \ z \ne z^*. \label{eq:pi_ijexample1a}
\end{eqnarray}

\noindent For $i \in \Omega_h, i^* \in \Omega_{h^*}$, $h, h^* (\ne h) = 1, \ldots, H$, and $z, z^* \in Z$,
\begin{equation}
\pi_{ii^*}(z, z^*) = \frac{r_h(z) r_{h^*}(z^*)}{N_h N_{h^*}}. \label{eq:pi_ijexample1b}
\end{equation}

\bigskip

\noindent \textbf{Assignment probabilities in Example 2:}

\medskip

\noindent For $i,i^* (\ne i) \in \Omega_h$ and $h = 1, \ldots, H$,
\begin{eqnarray}
\pi_i(z_1 z_2) &=& \frac{r_1(z_1) r_2(z_2)} { N}, \ \ \ z_1z_2 \in Z, \nonumber \\
\pi_{ii^*} (z_1z_2, z_1z_2) &=& \frac{r_1(z_1) r_2(z_2) \left(r_2(z_2)-1 \right)}{N(N_0-1)}, \ \ \ z_1z_2 \in Z, \nonumber \\
\pi_{ii^*} (z_1z_2, z_1z_2^*) &=& \frac{r_1(z_1) r_2(z_2) r_2(z_2^*)}{N(N_0-1)}, \ \ \ z_1z_2, z_1z_2^* (z_2 \ne z_2^*) \in Z, \nonumber \\
\pi_{ii^*} (z_1z_2, z_1^*z_2^*) &=& 0, \ \ \ z_1z_2, z_1^*z_2^* (z_1 \ne z_1^*) \in Z.
\label{eq:pi_ijexample2a}
\end{eqnarray}

\noindent For $i \in \Omega_h, i^* (\ne i) \in \Omega_{h^*}$ and $h, h^* (\ne h) = 1, \ldots, H$,
\begin{eqnarray}
\pi_{ii^*} (z_1z_2, z_1z_2^*) &=& \frac{r_1(z_1) \left( r_1(z_1)-1 \right) r_2(z_2) r_2(z_2^*) }{N N_0 (H-1)}, \ \ \ z_1z_2, z_1z_2^*  \in Z, \nonumber \\
\pi_{ii^*} (z_1z_2, z_1^*z_2^*) &=& \frac{r_1(z_1) r_1(z_1^*) r_2(z_2) r_2(z_2^*)}{N N_0 (H-1)}, \ \ \ z_1z_2, z_1^*z_2^* (z_1 \ne z_1^*) \in Z. \label{eq:pi_ijexample2b}
\end{eqnarray}

\bigskip

\noindent \textbf{Assignment probabilities in Example 3:}

\medskip

\noindent For $i, i^* (\ne i) \in \Delta(l)$ and $l=1, \ldots, |Z|$,
\begin{eqnarray}
\pi_i(z) &=& 1/|Z|, \ \ \ z \in Z, \nonumber \\
\pi_{ii^*}(z, z) &=& 1/ |Z|, \ \ \ z \in Z, \nonumber \\
\pi_{ii^*} (z, z^*) &=& 0, \ \ \  z, z^* (\ne z) \in Z, \label{eq:pi_ijexample3a}
\end{eqnarray}
while for $i \in \Delta(l), i^* \in \Delta(l^*)$ and $l, l^* (\ne l) = 1, \ldots, |Z|$,
\begin{eqnarray}
\pi_{ii^*}(z, z) &=& 0, \ \ z \in Z, \ \ \ \nonumber \\
\pi_{ii^*}(z, z^*) &=& \frac{1}{|Z|(|Z|-1)}, \ \ \ z, z^* (\ne z) \in Z. \label{eq:pi_ijexample3b}
\end{eqnarray}

\bigskip

\noindent \textbf{Proof of Theorem 2:}

\medskip

\noindent For any ${\bm Q} \in \mathbb{Q}$, from (\ref{eq:poplevelcontrast}), we have that
\small
\begin{eqnarray*}
\bar{\tau}^2 &=& \frac{1}{N^2}{\bm \tau}^{\prime} {\bm J} {\bm \tau} = {\bm \tau}^{\prime} {\bm Q} {\bm \tau} - {\bm \tau}^{\prime} \left( {\bm Q} - \left( 1/N^2 \right) {\bm J} \right){\bm \tau} \\
&=& {\bm \tau}^{\prime} {\bm Q} {\bm \tau} - \sum_{i=1}^N \sum_{i^*(\ne i) = 1}^N \left( q_{ii^*} -  1/N^2 \right) \tau_i \tau_{i^*}, \ \mbox{by the second equation in (\ref{eq:Q_cond})} \\
&=& {\bm \tau}^{\prime} {\bm Q} {\bm \tau} - \sum_{z \in Z} \sum_{z^* \in Z} \sum_{i=1}^N \sum_{i^*(\ne i) = 1}^N \left( q_{ii^*} -  1/N^2 \right) g(z) g(z^*) Y_i(z) Y_{i^*}(z^*), \ \mbox{by (\ref{eq:unitlevelcontrast})} \\
&=& {\bm \tau}^{\prime} {\bm Q} {\bm \tau} + \sum_{z \in Z} \sum_{z^* \in Z} \sum_{i=1}^N \sum_{i^*(\ne i) = 1}^N \left( M_{i i^*}(z, z^*) -  \widetilde{M}_{i i^*}(z, z^*) \right) Y_i(z) Y_{i^*}(z^*), \ \mbox{by (\ref{eq:M}) and (\ref{eq:tildeM})}.
\end{eqnarray*}
\normalsize
Substituting the expression above for $\bar{\tau}^2$ in (\ref{eq:thm1varest}) and applying (\ref{eq:upperbound}), we get
\begin{eqnarray*}
var(\taubarhat) = V_{\bm Q}(\taubarhat) - {\bm \tau}^{\prime} {\bm Q} {\bm \tau}.
\end{eqnarray*}

\noindent This proves part (1) of the Theorem. Part (2) follows from the fact that ${\bm Q}$ is psd by definition. To prove part (3), we note that because ${\bm Q}$ is psd, ${\bm \tau}^{\prime} {\bm Q} {\bm \tau}=0$ if and only if ${\bm Q} {\bm \tau} = 0$. From (\ref{eq:unitlevelcontrast}), ${\bm \tau} = \sum_{z \in Z} g(z) {\bm Y}(z)$. Thus, the upper bound in part (2) is attained for every treatment contrast if and only if ${\bm Q} {\bm Y}(z)$ is the same for each $z \in Z$, or equivalently, if and only if (\ref{eq:thm2cond}) holds.

\bigskip

\noindent \textbf{Proof of Proposition \ref{prop1}}

\medskip

\noindent By (\ref{eq:Q_cond}), ${\bm Q}$ has at least one eigenvalue equal to zero and trace $1/N$. Hence the mean of the positive eigenvalues of ${\bm Q}$ is at least $1/\left( N \left(N-1\right) \right)$. Because $\lambda_{\max}(\bm Q)$ is at least as large as this mean, it must satisfy
$$ \lambda_{\max}(\bm Q) \ge \frac{1}{N(N-1)}. $$
Equality is attained if and only if ${\bm Q}$ has $N-1$ eigenvalues equal to $1/\left( N \left(N-1\right) \right)$. There exists only one such ${\bm Q} \in \mathbb{Q}$, which is ${\bm Q}_{\text{strict}}$.

\bigskip

\noindent \textbf{Proof of Proposition \ref{prop2}}

\begin{itemize}
\item[(a)] By equations (\ref{eq:pi_ijexample2a}) and (\ref{eq:pi_ijexample2b}) in the Appendix, for the split-plot assignment mechanism, a psd matrix ${\bm Q} \in \mathbb{Q}$ satisfies (\ref{eq:SAP_suff}) if and only if  $q_{i i^*} = 1/N^2$, for every $i, i^* \in  \Omega_h$ and every $h = 1, \ldots, H$. Hence the if part is evident. To prove the only if part, observe that any psd matrix ${\bm Q} \in \mathbb{Q}$  can be expressed as ${\bm Q} = {\bm Q}_0 {\bm Q}_0^{\prime}$ for some ${\bm Q}_0$. Partition ${\bm Q}_0$ as ${\bm Q}_0 = [{\bm Q}^{\prime}_{01}, \ldots, {\bm Q}^{\prime}_{0H} ]^{\prime}$, where ${\bm Q}_{0h}$ has $N_0$ rows, $h=1, \ldots,H$. Suppose ${\bm Q}$ satisfies (\ref{eq:SAP_suff}). Then, as noted above, every diagonal block of ${\bm Q}$ has each element $1/N^2$, i.e., $ {\bm Q}_{0h} {\bm Q}_{0h}^{\prime} = (1/N^2) \left( {\bm 1}_0 {\bm 1}_0^{\prime} \right)$ for every $h$. Because the column space of ${\bm Q}_{0h}$ equals that of ${\bm Q}_{0h} {\bm Q}_{0h}^{\prime}$, it follows that, for every $h$, ${\bm Q}_{0h}$ has all elements equal in each column. Hence ${\bm Q} = {\bm Q}_0 {\bm Q}_0^{\prime}$  must be of the form ${\bm Q}_1 \bigotimes \left( {\bm 1}_0 {\bm 1}_0^{\prime} \right)$, for some square matrix ${\bm Q}_1$ of order $H$. The matrix ${\bm Q}_1$ has to be psd with each row sum zero and each diagonal element $1/N^2$, as ${\bm Q}$ is psd and ${\bm Q} \in \mathbb{Q}$.
\item[(b)] Any psd ${\bm Q}$ as in (a) has trace $1/N$ and at most $H-1$ positive eigenvalues. Hence the mean of the positive eigenvalues of ${\bm Q}$ is at least $1/ \left(N \left(H-1 \right) \right)$. The lower bound on $\lambda_{\max}({\bm Q})$ is now immediate as in Proposition \ref{prop1}. This bound is attained if and only if ${\bm Q}$ has $H-1$ eigenvalues equal to $1/ \left(N \left(H-1 \right) \right)$, i.e., if and only if ${\bm Q}_1$ has $H-1$ eigenvalues equal to $1/\left(N N_0 \left( H-1 \right) \right)$. As $N = HN_0$ and each row sum of ${\bm Q}_1$ is zero, this happens if and only if ${\bm Q}_1$ has each diagonal element $1/N^2$ and each off-diagonal element $-1/ \left(N^2 \left(H-1 \right) \right)$.	
\end{itemize}

\bibliographystyle{asa}
\small
\bibliography{MDRreferences}

\end{document}